\documentclass[lettersize,journal]{IEEEtran}
\usepackage{amsmath,amsfonts}
\usepackage{amsthm} %

\usepackage{array}
\usepackage[caption=false,font=normalsize,labelfont=sf,textfont=sf]{subfig}
\usepackage{textcomp}
\usepackage{stfloats}
\usepackage{url}
\usepackage{verbatim}
\usepackage{graphicx}
\usepackage{booktabs}
\usepackage{tikz}
\usetikzlibrary{positioning, arrows.meta, matrix}
\usepackage{cite}
\usepackage{hyperref}
\usepackage[capitalize,noabbrev]{cleveref}
\usepackage{siunitx}
\usepackage{orcidlink}
\newcommand\norm[1]{\lVert#1\rVert}

\usepackage{multirow} %
\usepackage{algorithm}
\usepackage{algpseudocode}
\usepackage{svg}
\usepackage{comment}
\usepackage{mathtools}
\usepackage{xfrac}

\newcommand{\D}[1]{\mathrm{d}{#1}}

\newcommand{\htime}{h_{\text{time}}}

\newcommand{\fse}{f_{\text{SE}}}
\newcommand{\Ldsm}{L_{\text{DSM}}}
\newcommand{\Ldp}{L_{\text{DP}}}
\usepackage{esvect}
\usepackage{adjustbox}
\usepackage{multirow}
\newcommand{\seqt}{\overrightarrow{t}}
\newcommand{\Vt}{\mathbf V^{\overrightarrow{t}}}
\newcommand{\Vtfk}{V^{\overrightarrow{t}}[\ell,k]}
\newcommand{\Tmax}{\text{T}_{\text{max}}}

\newcommand{\nfft}{\mathrm{n_{fft}}}

\usepackage[font=small,skip=0pt]{caption}
\usepackage[nolist, nohyperlinks]{acronym}

\usepackage{makecell} %

\begin{acronym}
\acro{mos}[MOS]{Mean Opinion Score}
\acro{bb}[BB]{Brownian Bridge}
\acro{db}[DB]{Diffusion Buffer}
\acro{rir}[RIR]{Room impulse responds}
\acro{nn}[NN]{Neural Network}
\acro{flop}[FLOPs]{Floating Point Operations per second}
\acro{dsm}[DSM]{Denoising Score Matching}
\acro{dp}[DP]{Data Prediction}
\acro{sgm}[SGM]{score-based generative model}
\acro{snr}[SNR]{signal-to-noise ratio}
\acro{gan}[GAN]{generative adversarial network}
\acro{vae}[VAE]{variational autoencoder}
\acro{ddpm}[DDPM]{denoising diffusion probabilistic model}
\acro{ncsnpp}[NCSN++]{Noise Conditional Score Network}
\acro{sncsnpp}[Sym-NCSN++]{Symmetric Noise Conditional Score Network}
\acro{stft}[STFT]{short-time Fourier transform}
\acro{istft}[iSTFT]{inverse short-time Fourier transform}
\acro{sde}[SDE]{stochastic differential equation}
\acro{bbed}[BBED]{Brownin Bridge with Exploding Diffusion Coefficient}
\acro{eum}[EuM]{Euler-Maruyama}
\acro{dint}[FL]{Frames-Lag}
\acro{ode}[ODE]{ordinary differential equation}
\acro{ou}[OU]{Ornstein-Uhlenbeck}
\acro{ve}[VE]{Variance Exploding}
\acro{dnn}[DNN]{deep neural network}
\acro{pesq}[PESQ]{Perceptual Evaluation of Speech Quality}
\acro{se}[SE]{speech enhancement}
\acro{bwe}[BWE]{bandwidth extension}
\acro{tf}[T-F]{time-frequency}
\acro{elbo}[ELBO]{evidence lower bound}
\acro{WPE}{weighted prediction error}
\acro{PSD}{power spectral density}
\acro{RIR}{room impulse response}
\acro{NN}{Neural Network}
\acro{mse}[MSE]{Mean-Squared-Error}
\acro{SNR}{signal-to-noise ratio}
\acro{LSTM}{long short-term memory}
\acro{POLQA}{Perceptual Objectve Listening Quality Analysis}
\acro{SDR}{signal-to-distortion ratio}
\acro{ESTOI}{Extended Short-Term Objective Intelligibility}
\acro{ELR}{early-to-late reverberation ratio}
\acro{TCN}{temporal convolutional network}
\acro{DRR}{direct-to-reverberant ratio}
\acro{nfe}[NFE]{number of function evaluations}
\acro{rtf}[RTF]{real-time factor}
\acro{SQA}{speech quality assessment}
\end{acronym}

\begin{document}

\title{Diffusion Buffer for Online Generative Speech Enhancement}

\author{Bunlong Lay\,{\orcidlink{0000-0002-0847-7896}}, %
Rostislav Makarov\,{\orcidlink{0009-0006-7319-6413}}, %
Simon Welker\,{\orcidlink{0000-0002-6349-8462}}, %
Maris Hillemann\,{\orcidlink{0009-0007-5701-8411}}, %
Timo Gerkmann\,{\orcidlink{0000-0002-8678-4699}},~\IEEEmembership{Senior Member,~IEEE}

\thanks{All authors are with the Signal Processing Group, Department of Informatics, Universität Hamburg, 22527 Hamburg Germany (e-mail: \{bunlong.lay; rostislav.makarov; simon.welker; timo.gerkmann\}@uni-hamburg.de).} 
}%

\markboth{Journal of \LaTeX\ Class Files,~Vol.~14, No.~8, August~2021}%
{Shell \MakeLowercase{\textit{et al.}}: A Sample Article Using IEEEtran.cls for IEEE Journals}

\maketitle

\begin{abstract}
Online Speech Enhancement was mainly reserved for predictive models. A key advantage of these models is that for an incoming signal frame from a stream of data, the model is called only once for enhancement. In contrast, generative Speech Enhancement models often require multiple calls, resulting in a computational complexity that is too high for many online speech enhancement applications. This work presents the \emph{Diffusion Buffer}, a generative diffusion-based Speech Enhancement model, which only requires one neural network call per incoming signal frame from a stream of data and performs enhancement in an online fashion on a consumer-grade GPU. The key idea of the Diffusion Buffer is to align physical time with Diffusion time-steps. The approach progressively denoises frames through physical time, where past frames have more noise removed. Consequently, an enhanced frame is output to the listener with a delay defined by the Diffusion Buffer, and the output frame has a corresponding look-ahead. %
In this work, we extend upon our previous work \cite{diffusionbuffer} by carefully designing a 2D convolutional UNet architecture that specifically aligns with the Diffusion Buffer's look-ahead. We observe that the proposed UNet improves performance, particularly when the algorithmic latency is low. Moreover, we show that using a Data Prediction loss instead of Denoising Score Matching loss enables flexible control over the trade-off between algorithmic latency and quality during inference. The extended Diffusion Buffer equipped with a novel neural network and loss function drastically reduces the algorithmic latency from \qty{320}{ms}--\qty{960}{ms} to \qty{32}{ms}--\qty{176}{ms} with an even increased performance. While it has been shown before that offline generative diffusion models outperform predictive approaches in unseen noisy speech data, we confirm that the online Diffusion Buffer also outperforms its predictive counterpart on unseen noisy speech data\footnote{Code will be released after acceptance.}.
\end{abstract}


\section{Introduction}
\IEEEPARstart{O}{nline} \ac{se} refers to enhancing a noisy speech signal on a frame-by-frame basis. This means that when one signal-frame arrives from a stream of noisy observations, the \ac{se} system needs to output one enhanced frame. In this work, we use the term \emph{online} when a small input-to-output delay up to \qty{200}{ms} is allowed. The ability to perform online \ac{se} in real-world scenarios is critical for ensuring clear and intelligible communication in video conferencing, VoIP calls, and other live interaction platforms. However, developing online \ac{se} systems is challenging. It requires low computational latency for processing a signal-frame, which makes the employment of computationally costly methods impractical. Therefore, online \ac{se} has to use a low-parameterized model that is capable of handling real-world data unseen during the development of these systems. %

Prominent online-capable Neural Networks \cite{defossez2020demucs, semamba} belong to the class of predictive (also known as discriminative) models. These approaches learn a direct mapping from the noisy observation to the clean target by training on pairs of clean and noisy mixtures in a supervised manner. During training, the different noise types, and a range of \acp{snr} are fixed. However, capturing all acoustic conditions during training that could occur in the real-world is not possible. As a result of this limitation, unpleasant distortions may occur when real-world data differ from training data.

Diverging from predictive approaches, generative approaches focus on learning a distribution of clean speech data, demonstrating promising results on unseen data \cite{richter_sgmse}. Recently, a category of generative models known as \emph{Diffusion models} has been introduced to the realm of SE \cite{lu2021study, lu2022conditional, welker2022speech, richter_sgmse, lay202interspeech}. The concept involves adding Gaussian noise to the data in the so-called \emph{forward process}, thereby transforming the data into a tractable distribution such as a Gaussian distribution. Subsequently, a \ac{nn} is trained to reverse this Diffusion process in a so-called \emph{reverse process} \cite{ho2020denoising}. Often, the forward and reverse processes are modeled by a forward and reverse \ac{sde}~\cite{richter_sgmse, jukic2024schr}. In the context of \ac{se}, the initial distribution is the clean speech data, and the terminating distribution is centered around the noisy mixture. Hence, these \acp{sde} can be thought of as a stochastic interpolation \cite{stochasticInterAlbergo} between the clean speech signal and the noisy mixture. Enhancement is then done by solving the reverse process, which is computationally intensive, requiring many evaluations of a large neural network. Therefore, the large computational burden of solving the reverse process makes Diffusion models impractical for online applications. 

In order to reduce the computational complexity of Diffusion models, many methods \cite{layCorrectingReverse, consistencymodel} aim to reduce the number of network evaluations. When approximating the so-called \ac{dsm} with a large \ac{nn}, called the score model, inference often requires a large number of score model evaluations for generating high-quality audio files. For instance, in \cite{richter_sgmse, lay202interspeech}, more than 30 score model evaluations are required for good performance. With 30 score model evaluations, the computational complexity is too large for many consumer-grade GPUs.
In \cite{stream_diffusion, lemercier2023storm}, the number of evaluations is reduced with the aid of a predictive model. However, still the complexity in \cite{stream_diffusion, lemercier2023storm} remains too high for many consumer-grade GPUs.

This paper expands our prior work, the \ac{db} \cite{diffusionbuffer}, which adapts SGMSE+ for streamable inference. The \ac{db} is the first generative \ac{se} method that achieves online speech enhancement on a consumer-grade GPU, an \emph{NVIDIA RTX 4080 Laptop} GPU, with a latency between \qty{320}{ms}--\qty{960}{ms}. A live demo was presented at the Interspeech 2025 conference \cite{lay_demo}, where code and video demonstrating its live performance can be found here\footnote{Video and code \url{https://github.com/sp-uhh/Diffusion-Buffer}}. The approach had been inspired by \cite{fifo, rollingdiff}, where the idea is to denoise frames through physical time. The \ac{db} is a buffer containing the most recent frames. Within this buffer, frames closer to the present are placed at a larger Diffusion time-step, while frames further in the past are progressively denoised. There are two important consequences of this idea. First, the $i$-th last output frame of the Diffusion Buffer has undergone $i$ reverse steps without the aid of any predictive guidance. In addition, it has a look-ahead of exactly $i-1$ frames. However, the symmetrical receptive field of the underlying 2D convolutional UNet architecture \ac{ncsnpp} does not align with this look-ahead constraint. In fact, whenever look-ahead frames are required but not available, then 
\ac{ncsnpp} used in \cite{diffusionbuffer} processes zeros instead of actual data. Therefore, suboptimal performance is achieved. Second, the Diffusion Buffer architecture achieves a drastic reduction in computational footprint as the underlying \ac{nn} is only required to be called once per incoming frame. However, as the underlying network is a large 2D  convolutional \ac{nn}, even one call may be too computationally heavy for consumer devices. For this reason, the parameters of that 2D convolutional \ac{nn} were reduced. It has been shown that Diffusion models perform very well when the underlying \ac{nn} has many parameters \cite{Podell2023SDXLIL}. However, it remains unclear if a lower-parameterized \ac{nn} still results in the good generalization to unseen data reported in \cite{richter_sgmse}. %

In addition, we expand the experimental framework of our previous publication \cite{diffusionbuffer} with the following contributions. We propose to change the formerly employed \ac{dsm} loss with the \ac{dp} loss, which achieves improved performance for lower latencies, e.g., when the algorithmic latency is only \qty{192}{ms} opposed to latencies between \qty{320}{ms}--\qty{960}{ms}. In addition, as employing the \ac{dp} loss retrieves a clean speech estimate with every \ac{nn} evaluation, any frame from the output of the Diffusion Buffer can be outputted to the listener.
Hence, employing the \ac{dp} loss allows for trading performance for latency during inference. This was not possible in the original implementation of the Diffusion Buffer \cite{diffusionbuffer}, which had a fixed delay during inference. In addition, we develop a 2D convolutional UNet where the receptive field aligns precisely with the look-ahead constraint of the Diffusion Buffer. The newly developed Diffusion Buffer with different loss and different underlying \ac{nn} outperforms largely the initial proposal in \cite{diffusionbuffer}. The proposed changes not only yield lower algorithmic latency between \qty{32}{ms}--\qty{192}{ms} but also lower computational processing time. In fact, with the proposed improvements, the Diffusion Buffer also runs online on a lower-end GPU (\emph{NVIDIA RTX 2080Ti}). Moreover, we revalidate the results of \cite{richter_sgmse, stream_diffusion}, as we find that the newly developed generative Diffusion Buffer performs better with unseen data than its predictive counterpart, i. e. the underlying architecture trained in a predictive manner.

\section{Background} \label{sec:background}
\subsection{Notation} With lowercase letters, we denote a vector with $v \in \mathbb{R}^n$, where $n>0$. In the context of \ac{se}, $v$ can be understood as a time-discrete signal and $n$ is the number of samples. In this case, their corresponding uppercase bold versions are complex-valued \ac{stft} spectra. More precisely, $\mathbf{V} \in \mathbb{C}^{F \times K}$ with $F$ being the number of frequency bins and $K$ the number of frames. We refer to an entry of $v$  with $v[i] \in \mathbb{R}$, $1\leq i \leq n$. Sometimes, we also write $v_i$ to refer to the $i-$th entry of a vector. For a single coefficient of a time-frequency bin of $\mathbf{V}$, we write $V[\ell,k] \in \mathbb{C}$, $1\leq \ell \leq F$, $1\leq k \leq K$. By abuse of notation, we are often only interested in the frames of the \ac{stft} $\mathbf V$, for which we write $\mathbf V[k] \coloneqq \mathbf V[\cdot,k] \in  \mathbb{C}^F$. Often, we are interested in a sequence of entries in $v$ starting from $i$ to $j$, $i<j$. In this case, we write $v[i:j] = [v[i], \dots, v[j]]^T \in \mathbb{R}^{j-i}$. In addition, we denote the hop-size, number of frequency bins, and frequency sampling rate of the \ac{stft} by $h_s$, $\nfft$, and $f_s$.

\subsection{Speech Enhancement}
\ac{se} is the task of retrieving the clean speech signal $s \in \mathbb{R}^n$ from a noisy observation $y \in \mathbb{R}^n $. In  general, we can write 
\begin{equation}
    y = u(s),
\end{equation}
where $u(\cdot)$ is the corruption operator. The precise definition of $u$ depends on which corruption type is considered. In this work, we only investigate on the additive corruption type that can be written as $u(s) = s + e$, where $e$ is environmental noise. However, applying to other corruption types is straightforward.

An \ac{se} model $\fse$ is a function that maps an input sequence $y = (y_1, \dots, y_n)^T$ to an output sequence $\hat{s} = (\hat{s}(1), \dots, \hat{s}(n))^T$, where $\hat{s}, y \in \mathbb{R}^n$. If $\fse$ is a well-parameterized \ac{se} model, then $\hat{s}$ is a good approximation of $s$. To perform enhancement, we distinguish between online and offline enhancement.

\subsection{Offline Inference} \label{sec:offline}
Offline processing (a.k.a. batch processing or utterance-based processing) is the evaluation of $\fse(y)$ by providing the whole sequence $y$ to the model $\fse(\cdot)$ at once. This assumes that the whole noisy observation $y$ is already available before the model begins to process any input. Therefore, this processing scheme cannot be used for enhancing a stream of audio data.%

\subsection{Online Processing} \label{sec:online}
Online processing (a.k.a. frame-by-frame processing) can be used to enhance a stream of audio data. For online enhancement in the \ac{stft} domain, we assume that in the $i-$th iteration, a new frame $\mathbf Y[i]$ arrives from the stream of noisy observations. More precisely, every hop-size $h_s$ we take the last $\nfft$ samples to transform them into a frame in the \ac{stft} domain. The model is evaluated on the new frame $\mathbf Y[i]$ together with all available past frames. The exact amount of past information depends on the width of the receptive field of the model. For \ac{se}, we believe that only a few hundred milliseconds of past information are sufficient to achieve good performance. 
Hence, limiting past information, opposed to all available past information, is reasonable and can be achieved with chunk-based processing.

A generic chunk-based processing scheme is described in \cref{alg:online1} where the present frame and past frames are simply stored in the current chunk $\mathbf Y_c$. This current chunk contains the last $M>0$ frames and is fed to the \ac{se} model $\fse$. From the evaluation $\mathbf O = \fse(\mathbf Y_c)$, the $d$-th last frame $O[M-d]$, $d \geq 0$, is then directed to the listener. Throughout this work, we call $d$ the \ac{dint} where a larger $d$ increases the algorithmic latency.

\subsubsection{Latency}
We define the algorithmic latency as the intrinsic latency of a processing system independent of its hardware constraints. Or differently stated, the algorithmic latency is the delay that would still exist even when processing on infinitely fast hardware. 
The algorithmic latency in seconds of \cref{alg:online1} is given by 
\begin{equation} \label{eq:alg_latency:stft}
\frac{\nfft}{f_s} + \frac{d \cdot h_s}{f_s}.
\end{equation}

For minimal algorithmic latency with such a scheme, it is desired to have $d=0$. The \ac{dint} parameter $d$ trades algorithmic latency for how much maximal look-ahead is possible in the enhanced output. Therefore, increasing $d$ may be reasonable when higher algorithmic latency is acceptable. 

The total latency is defined as the delay between the input signal and the output signal of a speech enhancement system. It is dominated by the sum of algorithmic latency and $\htime$, where $\htime \coloneqq \frac{h_s}{f_s}$ is the duration of a single frame hop in seconds. The total latency also includes other delays, such as the latency introduced by analog-to-digital conversion. We will not consider these other latencies in this work, as they are negligible. We therefore define the total latency as the sum of algorithmic latency and $\htime$.

\subsubsection{Real-time-factor}
An important constraint in \cref{alg:online1} is that the processing time $\text{proc-time}(\fse(\mathbf Y_c))$ must be shorter than $\htime$. We define the \ac{rtf} for streaming as 
\begin{equation} \label{eq:rtf}
    \text{RTF} = \frac{\text{proc-time}(\fse(\mathbf Y_c))}{\htime}
\end{equation}
By this defintion, we have that online \ac{se} is only possible if the \ac{rtf} is smaller than 1. If \ac{rtf} $\geq 1$, then processing an input frame is not completed before the next one arrives from the stream of audio data. This accumulates additional delays with each incoming frame and therefore prevents the \ac{se} system from operating online on an audio stream.

\begin{algorithm}
\caption{Chunk-based Online Speech Enhancement}\label{alg:online1}
\begin{algorithmic}[1]
\Require SE model $\fse(\cdot)$, $\mathbf Y$ stream of audio data in STFT.
\For{i = 0, 1, \dots}
    \State $\mathbf Y_c \gets \mathbf Y[i - M: i]$ \Comment{get last M frames}
    \State $\mathbf O = \fse(\mathbf Y_c)$ \Comment{compute estimate}
    \State output $\mathbf O[M-d]$ to the listener
\EndFor
\end{algorithmic}
\end{algorithm}

\section{Diffusion Models} \label{sec:diffusion_model}
\subsection{Stochastic Differential Equations} \label{sec:sde}
We formulate our Diffusion process in the \ac{stft} domain. Moreover, the following applies simultaneously to all time-frequency bins of $Y[\ell,k]$ of $\bold Y$. This means that the following equations are one-dimensional. For readability, we will omit indices $\ell,k$ in this section and reintroduce them when discussing loss functions from \cref{sec:db:dsm} on. Following the approach in \cite{lay202interspeech, richter_sgmse}, we model the forward process of the score-based generative model with an \ac{sde} defined on $0 \leq t < T_{\text{max}}$:
\begin{equation} \label{eq:fsde}
    \D{X_t} =
       f(X_t, Y) \D{t}
        + g(t)\D{{ w}},
\end{equation}
where $w$ is the standard Wiener process \cite{kara_and_shreve}, $X_t$ is the current process state with initial condition $X_0 =  S$, and $t$ is a continuous Diffusion time-step variable that describes the progress of the process that ends in the last Diffusion time-step $T_{\text{max}}$.
The term $f( X_t, Y) \D{t}$ can be integrated by Lebesgue integration \cite{rudin}, and $g(t)\D{{w}}$ follows Ito integration \cite{kara_and_shreve}. 
The diffusion coefficient $g$ regulates the amount of Gaussian noise that is added to the process, and the drift coefficient $f$ mainly affects the mean of $X_t$ (see \cite[(6.10)]{kara_and_shreve}) in the case of linear SDEs. The process state $X_t$ follows a Gaussian distribution \cite[Ch. 5]{sarkka2019sde}, called the \emph{perturbation kernel}:
\begin{equation}
\label{eq:perturbation-kernel}
    p_{0t}( X_t|X_0,  Y) = \mathcal{N}_\mathbb{C}\left(X_t; \mu_t(X_0, Y), \sigma_t^2 {I}\right).
\end{equation}
We call $\mu_t(X_0, Y)$ the \textit{mean evolution} and $\sigma_t$ the \textit{variance evolution} as they describe how the mean and variance of the process state $X_t$ are evolving over the Diffusion time $t$. If we can find analytically closed-form solutions for the mean and variance evolution, then \eqref{eq:perturbation-kernel} allows us to efficiently compute the process state $\mathbf X_t$ for each $t$ by calculating
\begin{equation} \label{eq:eff_sampling1}
    X_t = \mu_t( X_0,   Y) + \sigma_t z,
\end{equation}
with $z \sim \mathcal N_{\mathbb{C}}(0,1)$. By Anderson \cite{anderson1982reverse}, each forward SDE as in \eqref{eq:fsde} can be associated to a reverse SDE:
\begin{equation}\label{eq:plug-in-reverse-sde}
    \D{  X_t} =
        \left[
            - f(  X_t,  Y) + g(t)^2  \nabla_{  X_t} \log p_t(  X_t| Y)
        \right] \D{t}
        + g(t)\D{\bar{ w}}\,,
\end{equation}
where
$\D{\bar{ w}}$ is a Wiener process going backwards in time. In particular, the reverse process starts at $t=T$ and ends at $t=0$. Here $T < T_{\text{max}}$ is a parameter that needs to be set for practical reasons, as the last Diffusion time-step $T_{\text{max}}$ is only reached in limit. 

In this work, we employ the \ac{bbed} SDE from \cite{lay202interspeech}. The mean evolution is given by
\begin{equation}
\label{eq:mean_bb}
     \mu_t(X_0, Y) = \left(1- t\right) X_0 + t  Y
    \,,
\end{equation}
and the variance is
\begin{align} \label{eq:bbed:var}
   \hspace{-0.65em} \sigma(t)^2 &=  (1-t)c\left[(r^{2t}-1+t) + \log(r^{2r^2})(1-t)E \right], \\
    E &= \text{Ei}\left[2(t-1)\log(r)\right] - \text{Ei}\left[-2\log(r)\right],
\end{align}
where $\text{Ei}[\cdot]$ denotes the exponential integral function \cite{bender78:AMM, gradshteyn2007}, and $c, r > 0$ are \ac{sde} specific real-valued parameters.

The drift and diffusion coefficients are

\begin{align} \label{eq:bb-drift}
     f( X_t,  Y) &= \frac{ Y- X_t}{1-t},
\end{align}
and 
\begin{equation} \label{eq:ouve-diffusion-repara}
    g(t) = \sqrt{c}r^t, 
\end{equation}

There are different reverse process strategies to obtain $X_0$ from a given $X_T$ depending on the training objective. 

\subsubsection{Denoising Score Matching} \label{sec:db:dsm}
One straightforward way is to simply solve the reverse SDE in \eqref{eq:plug-in-reverse-sde}. To this end, the so-called \emph{score function} $\nabla_{ \mathbf X_t} \log p_t( \mathbf X_t|  \mathbf Y)$ in \eqref{eq:plug-in-reverse-sde} is approximated by a \ac{NN} $\mathfrak s_\theta(\mathbf X_t,\mathbf  Y, t)$. More precisely, the \ac{dsm} loss $\Ldsm$ is defined as:

\begin{equation}\label{eq:dsm}
     \Ldsm = \mathbb{E}_{t,(\mathbf X_0,\mathbf Y), \mathbf Z, \mathbf X_t|(\mathbf X_0,\mathbf Y)} \left[
        \norm{\mathfrak s_\theta(\mathbf X_t, \mathbf Y, t) + \frac{\mathbf Z}{\sigma_t}}_2^2
    \right]\,.
\end{equation}

Assuming that $\mathfrak s_\theta$ is available, we can generate an estimate of the clean speech $\mathbf X_0$ from $\mathbf Y$ by solving the reverse SDE \eqref{eq:plug-in-reverse-sde} for instance, with the first-order solver \ac{eum}.

\subsubsection{Data Prediction}
Another way is to simply predict the clean data with the \ac{dp} loss $\Ldp$:
\begin{equation}\label{eq:dp}
    \Ldp = \mathbb{E}_{t,(\mathbf X_0,\mathbf Y), \mathbf Z, \mathbf X_t|(\mathbf X_0,\mathbf Y)} \left[
        \norm{\mathfrak s_\theta(\mathbf X_t, \mathbf Y, t) - \mathbf X_0}_2^2
    \right].
\end{equation}

To obtain $\mathbf X_0$ from the initial $\mathbf X_T$, we iteratively refine the estimate given by $ \mathfrak s_\theta(\mathbf X_t, \mathbf Y, t)$. Starting with $\mathbf X_T$, one reverse step of this reverse process is given by
\begin{equation} \label{eq:dp:eff_sampling1}
    \mathbf X_{t_{i-1}} = \mu_{t_{i-1}}(\hat{\mathbf X}_0, \mathbf Y) + \sigma_{t_{i-1}} \mathbf Z,
\end{equation}
where $\mathbf Z \sim \mathcal N_{\mathbb{C}}(0,1)$. 
The clean signal estimate $\hat{\mathbf X}_0$ used in the mean evolution is obtained from $X_{t_i}$ by computing $s_{\theta}(\mathbf X_t, \mathbf Y, t)$.

\subsection{Latency considerations for streaming data} \label{sec:latency_offline_dm}

For the rest of this work, \cite{lay202interspeech, richter_sgmse} are referred to as offline Diffusion Models. To perform online enhancement with offline Diffusion Models as described in this section, we must call a large \ac{nn} many times in line 3 of \cref{alg:online1}. If we called the \ac{NN} only once, then usually the performance under the \ac{dsm} is of poor quality. When the \ac{NN} is trained with the \ac{dp}, then $N=1$ may yield reasonable quality, but does not differ much from a predictive method. An advantage of a generative model can be only observed when $N>1$, which we will show in \cref{sec:res:pred_vs_geb}.
However, calling the \ac{nn} multiple times is computationally not feasible with large \ac{nn}s. Therefore, we want to have only one \ac{nn} call in line 3 of \cref{alg:online1}, while still applying multiple reverse steps to each frame. The \ac{db} solves this problem at the cost of increased algorithmic latency.

\section{Diffusion Buffer} \label{sec:diff_buff}
In this section, we introduce the \emph{Diffusion Buffer} for \ac{se} from \cite{diffusionbuffer}, which originally optimizes on the \ac{dsm} in \eqref{eq:db:dsm} loss in \cite{diffusionbuffer}. We also propose to optimize on the \ac{dp} loss in \eqref{eq:db:dp}. In \eqref{eq:lookahead} we formulate the look-ahead constraint of the Diffusion Buffer for which in \cref{sec:arc_desgin} we design a 2D convolutional UNet whose receptive field aligns with the look-ahead constraint.

The work of \cite{diffusionbuffer} was inspired by \cite{rollingdiff, fifo}, where it was proposed to align the Diffusion time-steps with the time axis of the noisy mixture. To this end, we introduce a buffer, called Diffusion Buffer, containing the last $B$ frames, whereas the current frame $\mathbf R \in \mathbb{C}^{F \times 1}$ is placed at the end of this buffer and past frames are closer to the beginning of the buffer.
Within this buffer, frames that are closer to the end are modeled to be at larger Diffusion time-steps and therefore contain more noise. Equivalently, frames that are further in the past are progressively denoised until they become fully denoised. Consequently, with each new frame added into the buffer, we output a denoised frame that lies further in the past. Since there is a considerable delay between the current frame $R$ and the output frame, this approach increases algorithmic latency. 

When the training objective is the \ac{dsm} loss, then the added algorithmic latency is proportional to $B$, whereas when training the Diffusion Buffer on the \ac{dp} loss, then the additional algorithmic latency is proportional to \ac{dint} $d$. Whereas $d$, as before in \cref{alg:online1}, describes which output frame is taken to be directed to the listener. It can be flexibly adjusted during streaming to values between $1,\dots, B$. This differs from enhancing with the \ac{dsm} loss as the output frame is fixed to be the $(B-1)-$th last frame from the Diffusion Buffer. 

An important advantage of the Diffusion Buffer is that within the time of each hop length, the reverse process only requires a single evaluation of the \ac{nn}.

Mathematically, we formulate this process as follows: we fix $1\leq B \leq K$ as the number of reverse steps taken to enhance the frames of the noisy mixture.
Let $\seqt = (t_1, \dots, t_B)$ be an ascending sequence of Diffusion time-steps, i.e $0 < t_1 < \dots < t_B = T_{\text{max}}$. Then we call $\Vt \in \mathbb{C}^{F \times K}$ the \textit{perturbed input} and define it as

\begin{figure}
    \centering
    \includegraphics[width=1.0\linewidth]{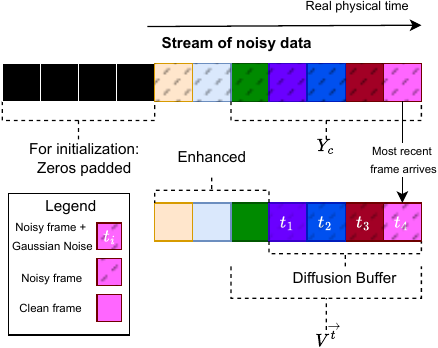}
    \caption{Diffusion Buffer illustration. Input to the \ac{NN} is the noisy chunk $Y_c$ and perturbed input $\Vt$, which contains the Diffusion Buffer. Frames within the Diffusion Buffer are modeled at different Diffusion-timesteps (white numbers in frame). In this example, $Y_c$ and $\Vt$ contain $K=5$ frames, and the Diffusion Buffer contains $B=4$ frames.}
    \label{fig:diffbuffer}
\end{figure}

\begin{equation} \label{eq:V}
\Vtfk \coloneqq
\begin{dcases*}
S[\ell,k]
   & if  $k < K-B$\,, \\[1ex]
X_{t_{\iota(k)}}[\ell, k]
   & if $k \geq K-B$\,.
\end{dcases*}
\end{equation}
where $\iota(k) = k - (K-B) + 1$ and $X_{t_{\iota(k)}}[\ell, k]$ can be computed via \eqref{eq:eff_sampling1}. In this formulation, the current frame from streamed data is placed at the last frame $V[\ell,K]^{\seqt}$. Moreover, we see that the frames $\Vtfk$ for $K-B\leq k \leq K$ are becoming more and more enhanced when close to $K-B$ and noisy when $k$ is close to $K$. All other frames outside of this buffer are already assumed to be cleaned. The collection of frames $\Vtfk$ with $K-B\leq k \leq K$ can also be thought of as a buffer on which we perform the diffusion process, hence the naming Diffusion Buffer.  \cref{fig:diffbuffer} shows an example of the Diffusion Buffer.

\subsection{Training} \label{sec:training}
For training, we fix the number of frames $K$, frequency bands $F$, the maximal number of reverse steps $B$ in the Diffusion Buffer, and the smallest Diffusion time-step $\epsilon > 0$.
We first sample a pair of clean and noisy files from the dataset. Then we pad the clean and noisy files with $K-1$ leading zeros in the STFT domain to mimic initialization when processing streamed data. In addition, we randomly crop a segment of $K$ frames from the clean and noisy file, which we call $S=X_0$ and $Y$, respectively, for the sequel. 
Second, we uniformly randomly sample an ascending sequence $\seqt = (t_1, \dots, t_B)$ of Diffusion time-steps with $t_1 = \epsilon > 0$. Third, based on \eqref{eq:eff_sampling1}, we compute for $K-B \leq k \leq K$
\begin{equation} \label{eq:eff_sampling2}
    X_{t_{g(k)}}[\ell, k] = \mu_{t_{g(k)}}(X_0[\ell,k], Y[\ell,k]) +\sigma_{t_{g(k)}} z(\ell,k)
\end{equation}
with $z(\ell,k) \sim \mathcal N_{\mathbb{C}}( 0, 1)$. We arrange $z(\ell,k)$ in a matrix as $\mathbf Z \in \mathbb{C}^{F\times B}$, likewise $\mathbf \Sigma \in \mathbb{C}^{F\times B}$. Fourth, we use \eqref{eq:eff_sampling2} to calculate $\Vtfk$ as in \eqref{eq:V} for all frequencies $1\leq \ell \leq F$ and frames $1\leq k \leq K$. Last, we can either optimize on the \ac{dsm} loss or on the \ac{dp} loss. The \ac{dsm} loss for the Diffusion Buffer is defined as:

\begin{equation}\label{eq:db:dsm}
      \Ldsm = \mathbb{E}_{t,(\mathbf X_0,\mathbf Y), \mathbf Z, \mathbf X_t|(\mathbf X_0,\mathbf Y)} \left[
        \norm{\mathfrak s_\theta(\mathbf \Vt, \mathbf Y, t) + \frac{\mathbf Z}{\mathbf \Sigma}}_2^2
    \right]
\end{equation}
where the division of matrices $\frac{\mathbf Z}{\mathbf \Sigma}$ is meant to be elementwise. The \ac{dp} loss for the Diffusion Buffer is given by:

\begin{equation}\label{eq:db:dp}
     \Ldp =  \mathbb{E}_{t,(\mathbf X_0,\mathbf Y), \mathbf Z, \mathbf X_t|(\mathbf X_0,\mathbf Y)} \left[
        \norm{\mathfrak s_\theta(\mathbf \Vt, \mathbf Y, t) - \mathbf A}_2^2
    \right],
\end{equation}
where $\mathbf A = \mathbf X_0[\cdot, K-B:B] \in  \mathbb{C}^{F\times K}$ is the clean target restricted to the last $K$ frames. Note that the input $\mathbf \Vt, \mathbf Y \in \mathbb{C}^{F\times K}$, but output of the network $\mathfrak s_\theta$ is in $\mathbb{C}^{F \times B}$.

\begin{algorithm}
\caption{Chunk-based processing for Diffusion Buffer}\label{alg:db}
\begin{algorithmic}[1]
\Require Reverse process $u_\theta$ with trained model $\mathfrak s_\theta$, noisy stream $\mathbf Y_s$, chunk size $K$, fixed Diffusion time-steps $\seqt = (t_1, \dots, t_B)$
\State $\hat{\mathbf S}$ $\gets$ [] \Comment{initialize output}
\State $\Vt$ $\gets [0, \dots, 0]$ \Comment{$K$ empty frames}
\For{frame $\mathbf R$ in $Y_s$}
    \State $\mathbf Y_c \gets$ last $K$ received frames \Comment{initialize with 0}
    \State $\Vt$ pops  \Comment{removes its first frame}
    \State $\Vt$ appends $ \mathbf R + \sigma_{t_B}\mathbf  Z$ \Comment{$\mathbf Z\sim \mathcal N_{\mathbb{C}}(\mathbf 0, \mathbf I_{F \times 1})$}
    \State $\mathbf O \gets \mathfrak s_\theta(\Vt, \mathbf Y_c, \seqt)$   \Comment{only one $\mathfrak s_\theta$ call}

    \State $\Vt$ $\gets u_\theta(O, \Vt, \mathbf Y_c, \seqt)$   \Comment{take one reverse step}
    \If{trained on $\Ldsm$}
        \State $\hat{\mathbf S}$ appends $(B-1)$-th last frame of $\Vt$
     \ElsIf{trained on $\Ldp$}
        \State $\hat{\mathbf S}$ appends $d$-th last frame of $O$
     \EndIf
\EndFor
\State \textbf{Return} $\Hat{S}$
\end{algorithmic}
\end{algorithm}

\subsection{Online inference for streamed data}
Once we have a trained model as described in \cref{sec:training}, we run inference by chunk-based processing as described earlier in \cref{alg:online1}. We adapt \cref{alg:online1} for the Diffusion Buffer in \cref{alg:db}. Let $\mathbf Y_s$ be an infinite stream of data in the STFT domain and assume we receive the frame $\mathbf R$ in the for-loop of \cref{alg:db}. We then add in line 6 an amount of Gaussian noise so that the random variable $\mathbf R' = \mathbf R + \sigma_{t_B} \mathbf Z$ follows the perturbation kernel \eqref{eq:perturbation-kernel}, meaning $\mathbf R'$ is at Diffusion time-step $t_B$. As usual, with the aid of the underlying trained \ac{nn}, we then run one reverse step for $\mathbf R'$ where the reverse process is either based on \ac{eum} (if trained on $\Ldsm$) or \eqref{eq:dp:eff_sampling1} (if trained on $\Ldp$). Consequently, $\mathbf R'$ is now at Diffusion time-step $t_{B-1}$. In fact, we run the reverse step for all frames within the \ac{db}. In particular, this means that the $(B-1)$-th last frame, which was before the reverse step at Diffusion time-step $t_1 = \epsilon$ is now at Diffusion time-step $0$. We therefore have enhanced this frame and output it to the listener when training on the \ac{dsm} loss \eqref{eq:db:dsm}. When training on the \ac{dp} loss \eqref{eq:db:dp}, then it is also possible to output the $d$-th last frame in line 12 of \cref{alg:online1} as an estimate of clean speech is available with each evaluation of $\mathfrak s_\theta$. An advantage of the \ac{dp} loss over the \ac{dsm} loss with the Diffusion Buffer is that the output frame can be flexibly adjusted during inference, as opposed to outputting the $(B-1)$-the last frame, where $B$ is fixed.

The algorithmic latency as described in \cref{sec:background} is $\frac{\nfft}{f_s} + \frac{(B-1)\cdot h_s}{f_s}$ when optimizing the Diffusion Buffer on the \ac{dsm} loss. For the \ac{dp} loss, the algorithmic latency is $\frac{\nfft}{h_s} + \frac{d \cdot h_s}{f_s}$.
Note that the reverse process only enhances frames within the \ac{db}, all other frames are not processed as they are already enhanced. As we advance with streamed data, we ensure that the output frame has undergone $B$ reverse steps when trained on \ac{dsm}, and $d+1$ reverse steps when trained on \ac{dp}. The zero part (black frames) in \cref{fig:diffbuffer} is required for initialization. For this, we intentionally padded the training data with leading zeros in the first step in \cref{sec:training} to match the initialization.

\subsection{Look-ahead Constraint of the Diffusion Buffer} \label{sec:look_constraing_db}
Recall from \cref{sec:background}, that we use the notation $\mathbf O[j]$ to refer to the frame $\mathbf O[\cdot ,j] \in \mathbb{C}^F$ as we shift our focus now to look-ahead constraints of the frames within the Diffusion Buffer.
An important aspect of the Diffusion Buffer is that the output frames $\mathbf O[i+K-B]$ depend on input frames $\mathbf Y[i+K-B:K]$
for $i=0, \dots, B$. In particular, the output frame
\begin{equation} \label{eq:lookahead}
\mathbf O[K-B+i] \text{ has a look-ahead of exactly } B-i \text{ frames}.
\end{equation}
We call \eqref{eq:lookahead}, the \emph{look-ahead constraint}. We hypothesize that using a \ac{NN} that satisfies this constraint is beneficial when outputting one of the last frames in line 12 of \cref{alg:db}. We believe the reason is that the last frames in the buffer process many uninformative zeros instead of only real data. As outputting one of the last frames relates to lower algorithmic latency (see \eqref{eq:alg_latency:stft}), we believe that we would achieve improved performance when algorithmic latency is low with such a \ac{NN}.
We will discuss in the next section why the \ac{NN} from \cite{diffusionbuffer} does not fulfill this constraint and how to design a \ac{NN} that meets the look-ahead constraint. In addition, we will see the benefits of such a \ac{nn} in \cref{sec:res}.

 \begin{figure*}[t]
     \includegraphics[width=\linewidth]{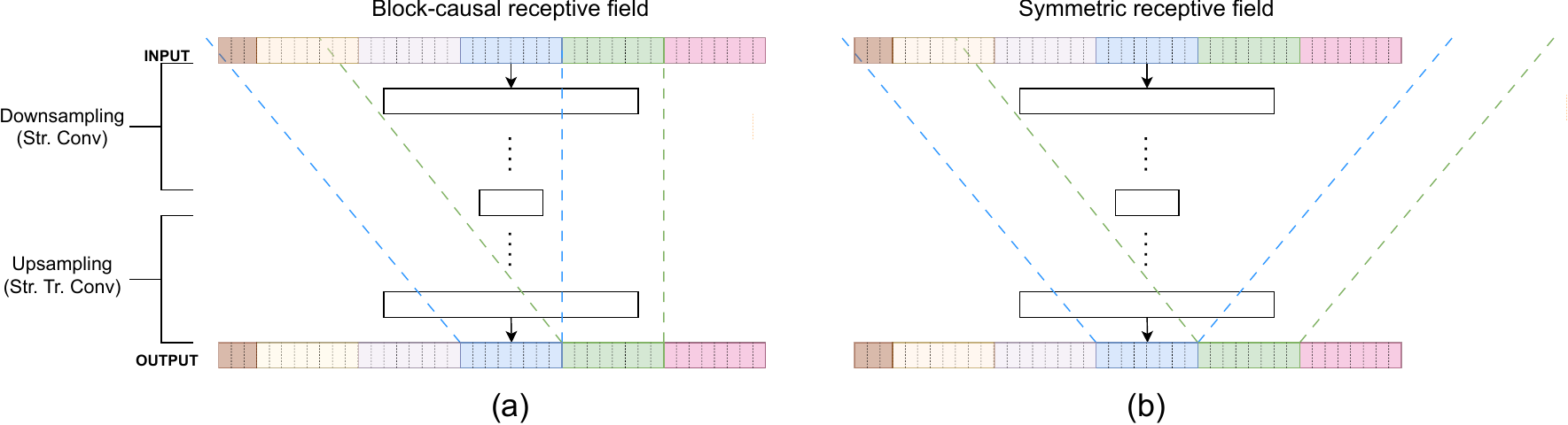}
     \caption{Two examples of receptive fields for an input sequence of length 43. The global stride of the network is 8. (a) Applying the three proposed modifications to the convolutional layers. The receptive field has block-causal dependencies. The look-ahead is at most equal to the global stride $g$. (b) Default down- and upsampling operations. The receptive is symmetrically placed around blocks and has a large look-ahead.}
     \label{fig:rec_field}
 \end{figure*}

\section{Architecture Design} \label{sec:arc_desgin}
We start our discussion by explaining \ac{ncsnpp} in \cref{sec:ncsnpp}, which serves as a starting point for our experiments. In addition, we show that the receptive field of \ac{ncsnpp} induces a large look-ahead in \cref{sec:ncsnpp_rf}. As a novelty, we provide precise instructions on how to modify \ac{ncsnpp} to fulfill the look-ahead constraint in \cref{sec:fulfill}, which can be applied to any UNet architecture based on convolutional layers.

In what follows, we consider an \ac{stft} representation as a sequence $\mathbf X$ of tokens $\mathbf X[i]$, where each token can be thought of as a frame of the \ac{stft}.

\subsection{NCSN++} \label{sec:ncsnpp}
The \ac{NN} \ac{ncsnpp} was originally introduced in \cite{song2021sde}. The first occurrence of the 2D UNet architecture \ac{ncsnpp} for \ac{se} is due to \cite{richter_sgmse}. In \cite{richter_sgmse}, \ac{ncsnpp} was adapted for the use on complex spectrograms by taking the real and imaginary parts of a complex spectrogram and using their concatenation as the network's input. The network is based on a multi-resolution U-Net structure where the downsampling and upsampling operations are realized by non-learnable finite impulse response (FIR) filters \cite{zhang2019making}. In addition, the network employs self-attention \cite{vaswani2017attention} and GroupNorm for normalization. Furthermore, the network incorporates a so-called progressive growing to stabilize high-resolution image generation \cite{karras2020analyzing}. 

\subsection{Receptive field of NCSN++} \label{sec:ncsnpp_rf}
There are several components that immediately affect the receptive field of \ac{ncsnpp}. First, there are self-attention mechanisms that are completely non-causal. Second, there is the GroupNorm that is also completely non-causal. When using either of these modules, the receptive field is at least the size of the input sequence. Therefore, we exclude these modules from the following discussion and focus on the remaining components' effects on the receptive field. 

Without self-attention and GroupNorm, a key aspect of determining the width of the receptive field of a 2D UNet architecture, such as \ac{ncsnpp}, are the down- and upsampling operations in that architecture. Since the down- and upsampling operations are done with symmetric padding, the width of the receptive field of \ac{ncsnpp} without self-attention and GroupNorm is also symmetric. This means in the context of \ac{se}, that if $r \in \mathbb{N}$ is the width of the receptive field, we need access to approximately $k-\frac{r}{2}$ input frames from the past and $k + \frac{r}{2}$ input frames from the future, in order to compute an $k$-th output frame $\mathbf O[k]$. In practice, the receptive field of \ac{ncsnpp} is relatively large. For instance, the receptive field of the used low-parameterized \ac{ncsnpp} with only 18M parameters from \cite{diffusionbuffer} without self-attention and GroupNorm is $r = 200$. This leads to a look-ahead of approximately $100$ tokens, which corresponds to approximately 1.6 seconds look-ahead in \cite{diffusionbuffer}. An illustration of a symmetric receptive field is given in \cref{fig:rec_field}b. In the case that the computation of an output token $\mathbf O[k]$ requires more future tokens than are available, zero-padding is used. This means that to compute the output token $\mathbf O[k]$, many zeros have been processed. This can result in performance degradation on the output token $\mathbf O[k]$ compared to actual data being processed instead of zeros.

\begin{figure}
    \centering
    \includegraphics[width=.9\linewidth]{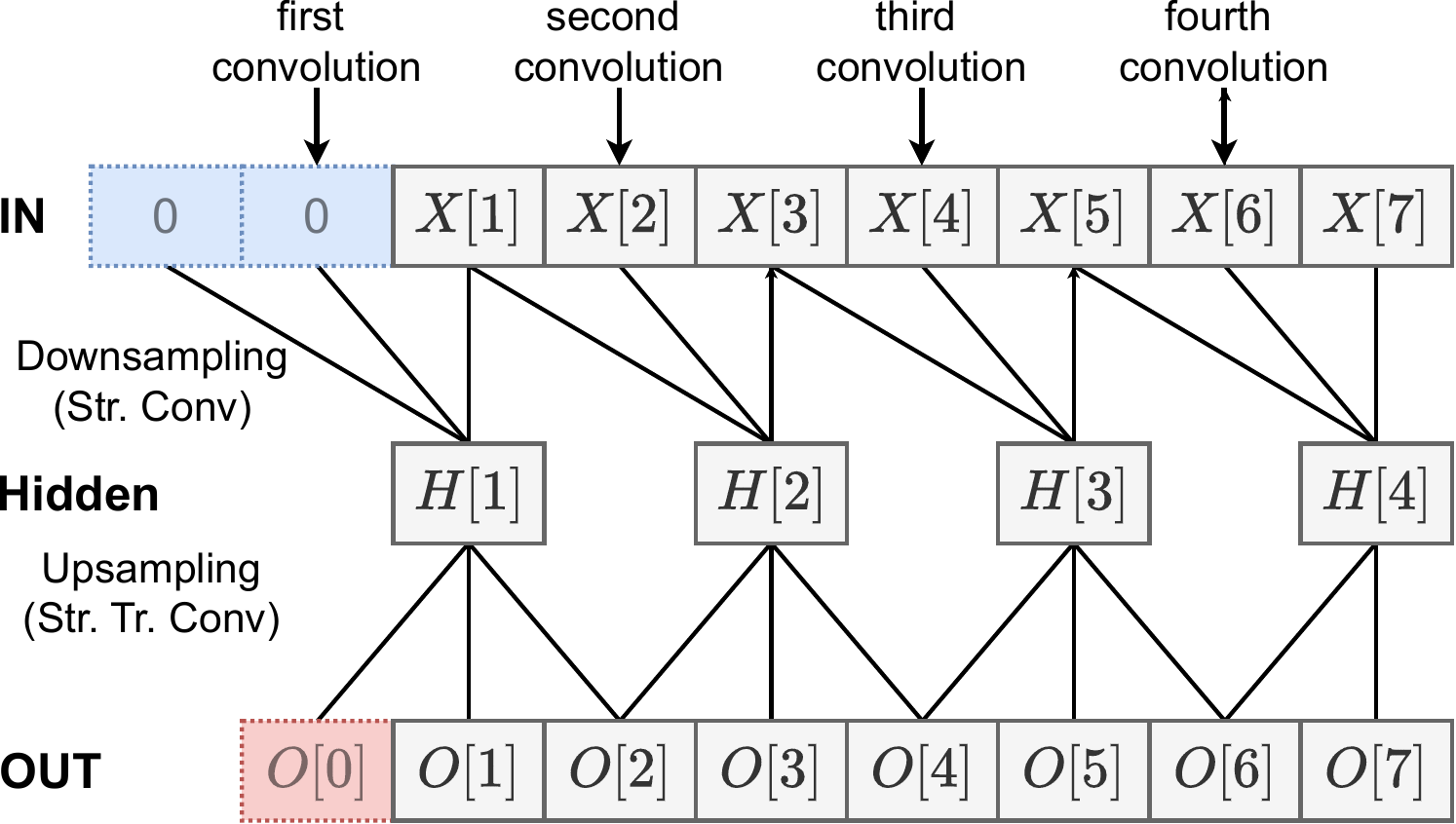}
    \caption{Proposed downsampling and upsampling on an input sequence $X$. Blue tokens represent zero-padding. The red token in the output will be cropped out, it will not be part of the final output.}
    \label{fig:down_up_exmp}
\end{figure}

\subsection{Modification of NCSN++} \label{sec:arc_desgin:ojsp}

As in \cite{ojsp}, we modify \ac{ncsnpp} as follows.
The downsampling and upsampling operations of \ac{ncsnpp} are done with non-learnable FIR filters. Instead of the FIR filters, we use 
learnable 2D convolutional layer and  2D transposed convolutional layer for the down- and upsampling operation with symmetrical padding, respectively. In addition, all attention layers were removed, and GroupNorm was replaced by cumulative GroupNorm.
Moreover, the progressive growing branch was also removed. As a consequence, the resulting network \ac{sncsnpp} has a symmetrical receptive field.

A causal attempt in \cite{ojsp} has been made by introducing causal 2D convolutional layers for the downsampling operation by padding and cropping the 2D convolutional layer of \ac{sncsnpp}. However, it has not been considered that the causality of the \ac{NN} changes when composing the downsampling with its corresponding upsampling operation when the stride is larger than 1. In this case, introducing causal 2D convolutional layers is insufficient to obtain a causal \ac{NN} that cascades several down- and upsampling operations. Instead of proposing only cropping and padding for the downsampling operation, we propose three design choices that achieve a \emph{block-causal} network. We will next see that a block-causal architecture automatically fulfills the look-ahead constraint described in \cref{sec:look_constraing_db}.

\subsection{Designing the Block-Causal Architecture} \label{sec:fulfill}

We start modifying the UNet architecture \ac{sncsnpp} to fulfill the look-ahead constraint. As a novelty, we make three important design choices that are different from standard convolutional layers:

\begin{itemize}
    \item Inference performance quality: We do not pad the input sequences of downsampling operations from the right with zeros.
    \item $p_\ell$-padding choice: We pad the input sequences of downsampling operations with exactly $p_\ell$ zeros from the left.
    \item Left-crop-upsampling: We only crop output sequences of upsampling operations from the left.
\end{itemize}

We motivate the inference performance quality design choice as follows. Assume that during inference with streaming data, sufficiently many past tokens are available. Then, since we do not zero-pad from the right in all downsampling layers, the newest output tokens are only affected by actual input data and do not process any zeros from zero-padding. We believe this design choice is important as processing zeros may degrade performance compared to processing actual data.
For the second design choice, let $k_\ell$ be the kernel-width, $s_\ell$ be the stride, and $n_\ell$ be the length of the input sequence to the $\ell-$th downsampling operation. The number of zeros padded from the left in the $\ell-$th downsampling layer is 
\begin{equation} \label{eq:pl}
    p_\ell = d_\ell s_k - n_\ell + (k_\ell-s_\ell),
\end{equation}
where $d_\ell= \lceil \frac{n_\ell}{s_\ell}\rceil$ is the maximum number of strides (partially) fitting into the input. %

The left-crop-upsampling choice is motivated by the fact that we padded with the $p_\ell$-padding choice from the left in its corresponding downsampling operation. The left padding in the downsampling operation results in some excess output length after upsampling, which we crop from the left.

In \cref{fig:down_up_exmp}, we see an example of an architecture following the design-choices. In this example, we have one downsampling and one upsampling operation with (non dilated) kernel-width $k=3$ and stride $s=2$. The figure shows that an input sequence $X$ of length $7$ is mapped to an output sequence $O$ of the same length (this length preservation holds for any input length). Note that for computing $H[4]$, no zeros were processed (inference performance quality). For computing $H[1]$ two zeros were processed ($p_\ell$-padding choice). For the upsampling operation, we center the kernel of the transposed strided convolutional layer around its input frame. In \cref{fig:down_up_exmp}, $H$ is the input of the upsampling operation and the output is given by $O$. Note that two effects may occur at the beginning and end of the output sequence of the upsampling operation. First, it may happen that the beginning of the output sequence, in this example $O[0]$, does not have a corresponding input frame $X[0]$. In this case, we simply crop the output sequence to the length of the input sequence $X$ (left-crop-upsampling). Second, it could happen that the end frame $O[8]$ cannot be computed yet as it requires more frames in the input, e.g., in \cref{fig:down_up_exmp} we would need $X[8], X[9]$ for computing $O[8]$. 

The result of executing the down- and upsampling operation on $X$ yields the following look-ahead. The last two output frames $O[6], O[7]$ depend on $X[6], X[7]$. In particular, $O[6]$ has a look-ahead of one frame and $O[7]$ does not have a look-ahead. For uneven $i$ we have that $O[i-1], O[i]$ depend on $X[i-1]$, $X[i]$, and $O[i-1]$ has a look-ahead of one frame and $O[i]$ does not have a look-ahead.

In \cref{fig:rec_field}a, we can see what happens if we stack multiple down- and upsampling operations obeying the three design-choices. To this end, we can assume that each downsampling operation has an arbitrary kernel width with a smaller stride. Let $g$ be the global stride, which we defined as the product of all strides in the downsampling operations. We then divide the input length $n$ of the sequence into blocks of length $g$. The output-to-input dependencies is best described in terms of blocks. The $i$-th output block depends up until the $i$-th input block. For this reason, we call such a model \emph{block-causal}. For instance, in \cref{fig:rec_field}a we see an example with $g=8$. In this example, we have that all output tokens of a color depend on their corresponding input tokens of the same color. For example, all blue output tokens depend on past tokens and on all blue input tokens, but have no dependencies on future tokens from the green and pink input tokens. In particular, we can describe the look-ahead behavior of such a model as follows. Let $O[i]$ be the $i-$th token of a block in the output, $i=1,\dots,g$, then

\begin{equation} \label{eq:fulfill_look}
O[i] \text{ has a look-ahead of exactly } g - i \text{ frames}.
\end{equation}

As we can see, this formulation matches the look-ahead constraint precisely, if we have that the global stride $g$ is equal to the buffer length $B$ of the Diffusion Buffer.

To summarize, to fulfill the look-ahead constraint of the Diffusion Buffer, we propose three design-choices. We do not pad from the right of the input sequence, we pad adequately with zeros from left in the downsampling operation and crop the output sequence of the upsampling operation to the same size as the input sequence of the downsampling operation. Cascading multiple down- and upsampling operations will yield a block-causal architecture. We report that we experimentally verified\footnote{Python script will be part of the repository upon acceptance} the block-causal receptive field of the \ac{nn} by backpropagating the gradients \cite{receptive_field}. We call the resulting block-causal architecture block-causal NCSN++ or briefly BC-\ac{ncsnpp}.

\subsection{Modification for Diffusion Buffer} \label{sec:arch_design:modification}
We derived BC-\ac{ncsnpp} from \ac{sncsnpp} by applying the three proposed design choices. However, for the \ac{db} we also need to modify the time-embedding layers as the Diffusion time-step sequence $\seqt$ is a $B$-dimensional vector, opposed to a scalar as it was before in \ac{ncsnpp}, \ac{sncsnpp}. To this end, we simply follow the modification of \cite{diffusionbuffer} made for the time-embeddings. We expand $\seqt$ with Fourier-embeddings to an $M \times B$ dimensional vector. Sequentially, for each down- and upsampling operation, the $M \times B$ dimensional vector is then transformed by a 2D convolutional layer to match the dimension of the down- or upsampling operation. The resulting time-embedding vector is then added to the down- or upsampling operation.

\section{Experimental setup} \label{sec:exp_setup}
\subsection{Data representation}  \label{sec:exp:data}
Each audio input, sampled at 16 kHz, is converted to a complex-valued \ac{stft}. As in \cite{diffusionbuffer}, we use a window size of $\nfft =510$ samples ($\approx \qty{32}{\ms}$), a hop length of $h_s=256$ samples ($\qty{16}{\ms}$), and a periodic Hann window. The input to the \ac{nn} for training is cropped randomly to $K=128$ time frames, resulting in approximately 2 seconds of data. A magnitude compression is used to compensate for the typically heavy-tailed distribution of \ac{stft} speech magnitudes~\cite{gerkmann2010empirical}. Each complex coefficient $v$ of the \ac{stft} representation is transformed as $\beta |v|^\alpha \mathrm e^{i \angle(v)}$, as in previous works \cite{diffusionbuffer, richter_sgmse} we use $\beta=0.5$, $\alpha=0.15$.

\subsection{Methods and Baselines} \label{sec:exp:methods}

\subsubsection{Diffusion Models} 
As mentioned in \cref{sec:diffusion_model}, we employ the BBED SDE from \cite{lay202interspeech} with $\Tmax = 0.999$, $k=2.6$ and $c=0.08$ following the parameterization from \cite{lay2024analysis}.
As in \cite{diffusionbuffer}, we used BBED for the Diffusion Buffer, which we refer to as DB-BBED. In addition, we can choose to optimize on the \eqref{eq:db:dsm} (online) or on the \eqref{eq:db:dp} (online). We will also experiment with different underlying architectures, whereas we either employ \ac{ncsnpp} or BC-\ac{ncsnpp}. When employing BC-\ac{ncsnpp}, we set the global stride $g$ to be the length of the Diffusion Buffer $B$ as this ensures that the look-ahead constraint is fulfilled as discussed in \eqref{eq:fulfill_look}.

\subsubsection{Predictive} We employ \ac{ncsnpp} or BC-\ac{ncsnpp} in a predictive setup. The input to the model is simply the degraded observation $y$ without the diffusion state and time-embeddings. This baseline is optimized on the \ac{mse} to predict the target directly. For a fair comparison, when we use these \ac{NN}s for the Diffusion models, we employ the same data representation as described in \cref{sec:exp:data}. The exact parameterization of both \ac{NN}s is in \cref{sec:NN_para}. 

In addition, we also train SEMamba \cite{semamba} in a causal way from the official repository. The model has 1.09G FLOPs and 6M parameters. For online enhancement, we use the official streaming pipeline from the repository \footnote{\url{https://github.com/RoyChao19477/SEMamba}}.

\begin{table} 
\begin{center}
\caption{Computational demands of BC-\ac{ncsnpp} and \ac{ncsnpp} when time-embeddings are removed. \ac{rtf} computed based on \eqref{eq:rtf} with \texttt{torch.compile} on two NVIDIA GPUs (RTX 4080 Laptop, RTX 2080 Ti) with one second chunk. FLOPs computed with \texttt{torch.utils.flop\_counter}.}
    \begin{tabular}{c|cccc|}
        NN & RTF & Num. Parameter & FLOPs \\
        & 2080Ti / 4080 & M & Giga \\
        \hline
        BC-NCSN++ & 0.97 / 0.44 &   15.35  & 56  \\
        $g=16$&  & & \\
        \hline
        BC-NCSN++ & 0.97 / 0.44 & 22.18 & 56   \\
        $g=32$&  & & \\
        \hline
        NCSN++  &  1.15 / 0.63 &  18.28 & 87  \\
        \hline
    \end{tabular}
\end{center}
     \label{tab:NN_para}
\end{table}

\subsection{Parameterization and Training of BC-NCSN++, NCSN++} \label{sec:NN_para}
We used the following parameterization for BC-NCSN++ and NCSN++ in \cref{tab:NN_para}. We mention first that the reported \ac{flop}, \ac{rtf}, and number of trainable parameters are based on evaluation without any time-embeddings, which are necessary for the Diffusion Buffer (see \cref{sec:arch_design:modification}). When running online inference with the Diffusion Buffer, we can simply cache the time-embedding operations as they do not depend on actual data. Therefore, we can disregard time-embeddings, and the numbers from \cref{tab:NN_para} remain the same when the Diffusion Buffer employs one of the \ac{NN} from \cref{tab:NN_para} during inference with cached time-embeddings.

The reported \ac{ncsnpp} from \cref{tab:NN_para} follows the parameterization from \cite{diffusionbuffer}. More precisely, the channel dimensions are $(96, 96, 96, 96, 96)$ with downsampling/upsampling factors $(1, 2, 2, 2)$. The BC-NCSN++ architecture with $g=16$ has channel dimension $(128, 256, 256, 256, 128)$ with downsampling/upsampling factors $(2, 2, 2, 2)$. For $g=32$, we double the last channel dimension to 256, and the last downsampling factor to 4. %

For training DB-BBED and the predictive methods using NCSN++ or BC-NCSN++, we optimized with ADAM \cite{kingma2015adam} with a learning rate of $0.0001$ with a batch-size of $32$. We track an exponential moving average of the DNN weights with a decay of 0.999. We trained a total of 200 epochs, which lasted for approximately 1 day on an NVIDIA A100.

\subsection{Datasets} \label{sec:exp:dataset}
We train, validate and test on the publicly available dataset EARS-WHAM-v2\footnote{\url{https://github.com/sp-uhh/ears\_benchmark}} with clean files from the EARS dataset \cite{richter2024ears} and noise files from the WHAM dataset \cite{wham}. The dataset, originally recorded at 48 kHz, was downsampled for this work to 16 kHz. This dataset has 54 hours for training, 1.1 hours of validation, and 3.5 hours for testing. 

To test on mismatched conditions, we generate a test set containing impulsive noise types. For this end, we take files from the SoundIdeasGeneral 6000 (www.sound-ideas.com) data set. We filter for classes that potentially contain impulsive noise files. Hence, we check if the class name of a file contains one of the following keywords: `ball`, `gun`, `footsteps`, `archery`, `drop`, and `tennis`. In addition, we use a heuristic method to detect if the file is indeed impulsive. For this, we check if the kurtosis of a time-domain file is larger than 10 \cite{ANTONI2006282}. Moreover, we check if the duration of the noise file is between 0.5 and 20 seconds, and if
less than 70 percent of the noise file is active.
Finally, we mixed the impulsive noise files with the clean files from the EARS-WHAM-v2 test set with an SNR between -7 to 5 dB at 16 kHz. We call the resulting test set of 1.5 hours duration EARS-General.

\subsection{Metrics}\label{sec:exp:metrics}
To evaluate the performance of the proposed method, we use standard metrics, which we will describe in detail below. Higher values indicate better performance for all metrics.%
\paragraph{PESQ} The Perceptual Evaluation of Speech Quality (PESQ) is used for objective speech quality testing and is standardized in ITU-T P.862 \cite{rixPerceptualEvaluationSpeech2001}. The PESQ score lies between 1 (poor) and 4.5 (excellent). We use the wideband PESQ version.

\paragraph{ESTOI} 
The Extended Short-Time Objective Intelligibility (ESTOI) is an instrumental measure for predicting the intelligibility of speech subjected to various kinds of degradation \cite{jensen2016algorithm}. The metric is normalized and lies between 0 and 1, with higher values indicating better intelligibility.

\paragraph{SI-SDR, SI-SIR} 
Scale-Invariant (SI-) Signal-to-Distortion Ratio (SDR), Signal-to-Interference Ratio (SIR) are standard evaluation metrics for single-channel speech enhancement and speech separation\cite{leroux2018sdr} which are measured in dB.

\paragraph{DistillMOS} DistillMOS \cite{stahl2025distillation} is a \ac{mos} prediction method, built by distilling a wav2vec2.0-based \ac{SQA} model into a more efficient model.

\begin{figure}
    \centering
\includegraphics[width=1.0\linewidth]{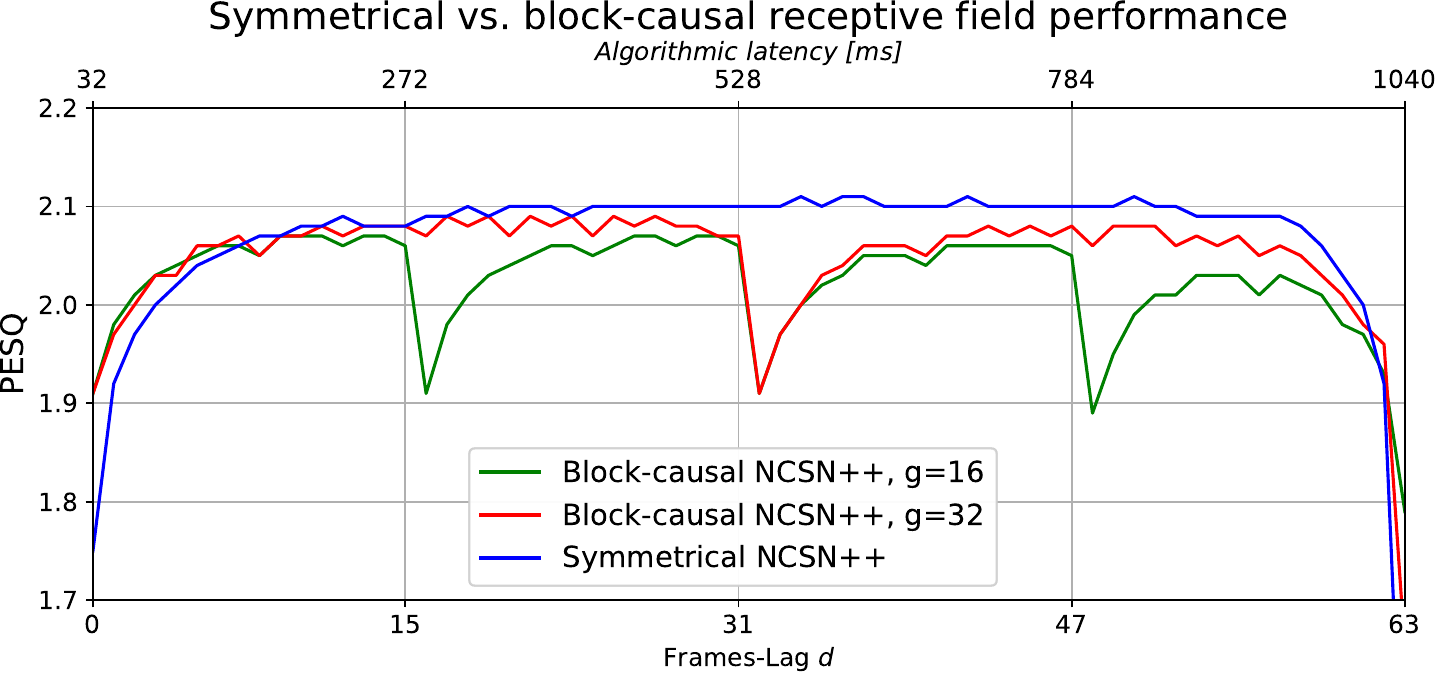}
    \caption{ Comparison of \ac{sncsnpp} (symmetric receptive field) versus BC-NCSN++ (block-causal receptive field) on EARS-WHAM-v2. Chunk-based (online) processing performance plotted against different \acf{dint} $d$. We see that \ac{sncsnpp} degrades much more for small $d$ opposed to BC-NCSN++.}
    \label{fig:bc_analysis}
\end{figure}

\section{Results} \label{sec:res}
The following results are based on online processing. For predictive methods \ac{ncsnpp}, \ac{sncsnpp}, and BC-\ac{ncsnpp}, we used the chunk-based algorithm as described in \cref{alg:online1}, and for the generative DB-BBED approaches, we used the chunk-based Diffusion Buffer adoption as described in \cref{alg:db}. Although these methods were trained on $K=128$ ($\approx$ 2 seconds) chunks, for inference we used only $K=64$ ($\approx$ 1 second), as lowering the time context to $K=64$ does not decrease performance.
SEMamba uses its streaming pipeline from its official repository.

In \cref{sec:res:bc_analysis}, we evaluate the performance of the proposed BC-\ac{ncsnpp} against its symmetrical version \ac{sncsnpp}. We will see that these methods achieve a similar maximal PESQ value, when the \ac{dint} $d$ is large enough. However, an advantage of the block-causal receptive field is that at low $d$, NC-\ac{ncsnpp} largely outperforms \ac{sncsnpp}, justifying the need of a block-causal over a symmetrical receptive field. 
This means BC-\ac{ncsnpp} has better performance in applications with low algorithmic latency, but achieves similar performance when algorithmic latency is allowed to be large.

In \cref{sec:res:ablation_db}, we present an ablation study of the Diffusion Buffer in terms of changing the loss function and the underlying \ac{nn} to BC-\ac{ncsnpp}.

Last, in \cref{sec:res:pred_vs_geb} we compare DB-BBED against its predictive counterpart. We observe that on the matched test set, the predictive counterpart has in most metrics a small advantage. However, we observe that on the unseen noise types in the mismatched test set, the generative methods now outperform the predictive method when we take more reverse steps. In fact, the predictive method is sometimes not able to remove the unseen impulsive noise types.

\subsection{Analysis of Block-Causal NCSN++} \label{sec:res:bc_analysis}
In \cref{fig:bc_analysis}, we see how the block-causal architecture BC-NCSN++ compares against its counterpart \ac{sncsnpp} with a symmetric receptive field when trained and tested on EARS-WHAM-v2 in a predictive manner as described in \cref{sec:exp:methods}. The parameterizations for $g=16, 32$ are described in \cref{sec:NN_para}. In addition, \ac{sncsnpp} from \cref{fig:bc_analysis} follows the same parameterization as for BC-NCSN++ with $g=16$. It is therefore directly comparable, as the symmetric and block-causal versions have the same \ac{nn} architecture, except for the padding and cropping in the down- and upsampling operations based on the design choices detailed in \cref{sec:fulfill}.%

We observe a periodic behavior of BC-NCSN++ with periodicity equal to the global stride $g$. The performance degrades to a minimum for every $d$-th output frame, which does not have any look-ahead. For instance, for BC-NCSN++ $g=16$ (green solid line), this is given by $d=0, 16, 32, 48$. When the output frame has a look-ahead, then performance also increases, whereas we observe that we have a diminishing return on performance when allowing for more look-ahead. For instance, performance barely increases when the look-ahead is more than 10 frames ($\geq \qty{144}{ms}$), as the green and red solid lines become almost constant in each periodic block.
Likewise, \ac{sncsnpp} (blue solid line) remains virtually constant in PESQ when $d\geq 15$ ($\geq \qty{240}{ms}$). These results hint that \ac{se} is a highly temporally local problem, which means that a look-ahead of $\qty{144}{ms}$ - $\qty{240}{ms}$ is already sufficient for optimal \ac{se} performance. A similar finding has been observed in \cite{exploretradeoffsWilson}, where $\qty{200}{ms}$ look-ahead is sufficient for optimal performance.%

Moreover, we observe that BC-NCSN++ ($g = 16, 32$) achieves a similar maximal PESQ value as \ac{sncsnpp} when the delay $d \mod g$ is large enough. Concretely, we find a maximum PESQ value of $2.11$ for \ac{sncsnpp} at $d=35$. For
BC-\ac{ncsnpp} with $g=16$ and $g=32$, the maximum PESQ values are $2.07$ and $2.09$ respectively, with also only a minor difference ($0.04$ PESQ).
Other metrics lead to similar conclusions, see \cref{sec:exp:metrics}. We therefore conclude that the block-causal symmetrical field given by BC-\ac{ncsnpp} achieves similar performance as its symmetrical receptive field counterpart in \ac{sncsnpp}.

However, we see an advantage of BC-NCSN++ over \ac{sncsnpp} when $d$ is small. In this case, \ac{sncsnpp} degrades quickly to a much lower performance than BC-NCSN++ with $g=16, 32$. This is simply explained by the fact that \ac{sncsnpp} processes many uninformative zeros from the symmetrical zero-padding used in the architecture. In contrast, due to the design-choices made in \cref{sec:fulfill}, BC-NCSN++ does not degrade as much as its symmetric counterpart \ac{sncsnpp}. For $d=0$, the difference between BC-NCSN++ with $g=16$ and \ac{sncsnpp} is more than $0.16$ in PESQ in favor of BC-NCSN++. This is a key advantage of the block-causal architecture over its symmetrical version \ac{sncsnpp}, as it allows for lowering the algorithmic latency without sacrificing too much performance.

\begin{figure}
    \centering
    \includegraphics[width=1.0\linewidth]{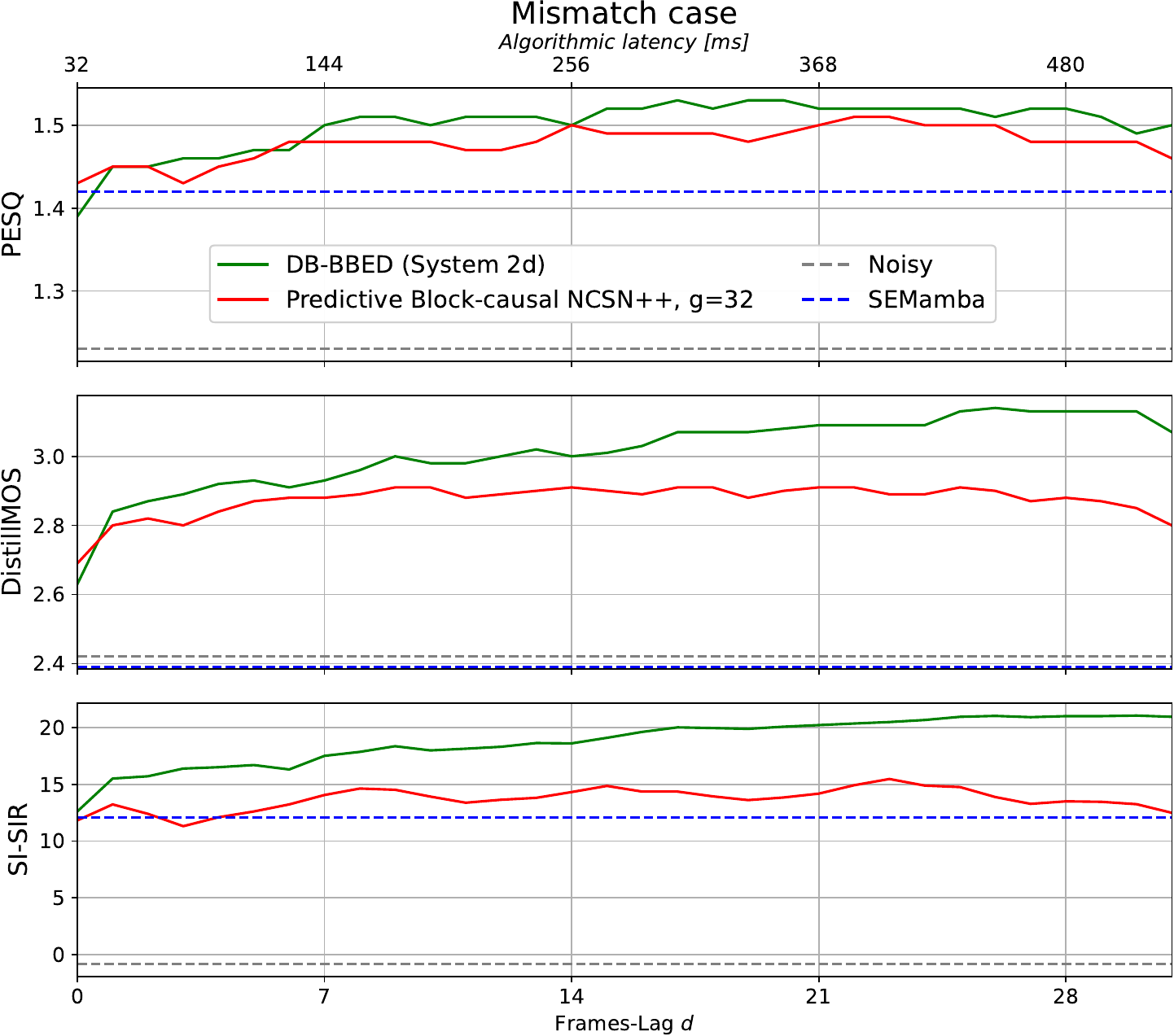}
    \caption{Comparison between generative DB-BBED and its predictive counterpart on the mismatched EARS-General test set. Unlike the matched case in \cref{tab:db_ablation}, we can now see that the generative method outperforms the predictive method.}
    \label{fig:mismatch_gen_vs_pred}
\end{figure}

\begin{table*}
    \centering
    \caption{Results on EARS-WHAM-v2 when noisy files were chunk-based processed online. $d$ is the \ac{dint} from which the alg. latency can be computed as described in \cref{sec:online}. $N$ is the number of reverse steps. $B$ for DB-BBED.}
    \label{tab:db_ablation}
    \setlength{\tabcolsep}{3pt} %
    \resizebox{\textwidth}{!}{ %
    \begin{tabular}{c|ccccccccc}
    \toprule
    Method & System & NN & $d$ / $N$ / Alg. latency & Loss & PESQ & ESTOI & DistillMOS & SI-SDR & SI-SIR  \\
    \midrule
    Noisy & - & - & - & - & $1.24 \pm 0.21$ & $0.64 \pm 0.17$ & $2.58 \pm 0.60$ & 
    $5.36 \pm 5.90$ & $5.36 \pm 5.90$  \\

    \midrule
    
    \multirow{5}{*}[-0.6ex]{\rotatebox[origin=c]{90}{Predictive}}  & 1a & NCSN++& 9 / - / 176 ms & MSE & 
    $2.26 \pm 0.62$ & $\mathbf{0.84 \pm 0.12}$ & $\mathbf{3.97 \pm 0.59}$ & 
    $\mathbf{15.36 \pm 6.32}$ & $\mathbf{28.77 \pm 5.18}$  \\
    \rule{0pt}{4ex}
     & 1b &\begin{tabular}{@{}c@{}}BC NCSN++\vspace{-.9mm}\\{\scriptsize{$g=32$}}\end{tabular}&  9 / - / 176 ms & MSE & 
    $2.06 \pm 0.59$ & $0.82 \pm 0.12$ & $3.93 \pm 0.62 $ & 
    $14.77 \pm 4.85$ &  $28.41 \pm 5.12$\\ 
    \rule{0pt}{4ex}
    
     &1c & SEMamba & - / - / 25 ms & see [3] & 
     $\mathbf{2.61 \pm 0.54}$ & $0.82 \pm 0.12$ & $3.75 \pm 0.58$ & 
     $8.84 \pm 3.77$ &  $27.70 \pm 5.46$ \\  
     
    \midrule
    
    \multirow{4}{*}[-4.5ex]{\rotatebox[origin=c]{90}{DB-BBED}}  & 2a & \begin{tabular}{@{}c@{}}NCSN++\vspace{-.9mm}\\{\scriptsize{$B=16$}}\end{tabular}  & 15 / 16 / 272 ms & DSM & 
    $1.59 \pm 0.37$ & $0.75 \pm 0.15$ & $3.70 \pm 0.72$ & 
    $12.16 \pm 5.41$ &  $23.35 \pm 5.91$ \\
    \rule{0pt}{4ex}
     &2b &\begin{tabular}{@{}c@{}}BC NCSN++\vspace{-.9mm}\\{\scriptsize{$g=16=B$}}\end{tabular} & 15 / 16 / 272 ms & DSM & 
     $1.64 \pm 0.43$ & $0.71 \pm 0.18$ & $3.59 \pm 0.81$ & 
     $10.97 \pm 5.56$ &  $23.62 \pm 6.61$ \\
    \rule{0pt}{4ex}
     & 2c &\begin{tabular}{@{}c@{}}BC NCSN++\vspace{-.9mm}\\{\scriptsize{$g=16=B$}}\end{tabular}   & 9 / 10 / 176 ms & DP & 
    $1.81 \pm 0.55$ & $0.77 \pm 0.17$ & $3.68 \pm 0.71$ & 
    $12.51 \pm 5.60$ & $24.71 \pm 6.70$  \\
    \rule{0pt}{4ex}
     &   2d & \begin{tabular}{@{}c@{}}BC NCSN++\vspace{-.9mm}\\{\scriptsize{$g=32=B$}}\end{tabular}  & 9 / 10 / 176 ms & DP & 
     $\mathbf{2.02 \pm 0.59}$ & $\mathbf{0.81 \pm 0.13}$ & $\mathbf{3.76 \pm 0.67}$ & 
     $\mathbf{14.35 \pm 5.14}$ & $\mathbf{25.42 \pm 6.35}$  \\
     
     \bottomrule
    \end{tabular}
    }
\end{table*}

\subsection{Ablation on Diffusion Buffer in matched conditions} \label{sec:res:ablation_db}
In \cref{tab:db_ablation}, we can see the results on the EARS-WHAM-v2 test set of the predictive methods BC-\ac{ncsnpp} ($g=16$) and \ac{ncsnpp}, whereas the three methods follow the parameterization described in \cref{sec:NN_para}. All methods are evaluated in a chunk-based online processing scheme. As we observed from \cref{sec:res:bc_analysis} that with $d=9$ performance remains stable, we set for the experiments $d=9$.
We observe that \ac{ncsnpp} outperforms BC-\ac{ncsnpp} and SEMamba in all reported metrics except for PESQ. As SEMamba is optimized on PESQ, it achieves a relatively large PESQ value, even with a low latency of $\qty{25}{\ms}$.

System 2a of \cref{tab:db_ablation} is the original Diffusion Buffer from \cite{diffusionbuffer} with $B=16$. As we can see, this performs poorly compared to the predictive methods. For System 2b, we changed \ac{ncsnpp} for BC-\ac{ncsnpp} with $g=16$. This system shows similar performance as when trained with \ac{ncsnpp}, although BC-\ac{ncsnpp} has lower computational demand in terms of number of parameters, \ac{rtf} and \ac{flop} than \ac{ncsnpp}, as can be seen from \cref{tab:NN_para}. An improvement in all metrics can be observed in System 2c, when changing the loss function to the \ac{dp} loss, albeit lowering \ac{dint} $d$ and therefore using fewer reverse steps. As for the predictive methods, we report that we do not observe performance improvement when increasing $d$ beyond $d=9$, we therefore set $d=9$. Note that changing \ac{dint} $d$ from $15$ to $9$ also decreases the algorithmic latency by $\qty{96}{\ms}$ down to $\qty{176}{\ms}$. Last, we changed BC-\ac{ncsnpp} with $g=16$ to BC-\ac{ncsnpp} with $g=32$, which again improves the Diffusion Buffer (compare System 2c versus 2d). The improvement is possibly due to an increase in the \ac{nn}'s number of parameters. This behavior differs from the predictive counterparts, as discussed in \cref{sec:res:bc_analysis} where both predictive counterparts achieve similar maximal PESQ value. The newly developed Diffusion Buffer (System 2d) with different loss and different \ac{nn} largely outperforms the first implementation of Diffusion Buffer (System 2a) from \cite{diffusionbuffer} in all reported metrics. System 2d even reduces the algorithmic latency by $\qty{96}{\ms}$, and uses a \ac{nn} that even performs online on an NVIDIA 2080 Ti, opposed to the employed \ac{ncsnpp} of System 2a (see \cref{tab:NN_para}).

\subsection{Predictive versus Generative} \label{sec:res:pred_vs_geb}
Comparing the generative DB-BBED in System 2d from \cref{tab:db_ablation} with its predictive counterpart in System 1b, we can see that the predictive counterpart marginally outperforms the generative DB-BBED in most metrics.
In DistillMOS we observe a relatively large gap of $0.17$. In addition, we report that when $d=0$, or equivalently \qty{32}{ms} latency, System 2d performs worse than SEMamba.

In \cref{fig:mismatch_gen_vs_pred}, we compare the generative DB-BBED (System 2d from \cref{tab:db_ablation}) against its predictive counterpart (System 1b from \cref{tab:db_ablation}). Results are based on testing on the mismatched EARS-General test set. 
In the matched case in \cref{tab:db_ablation}, generative DB-BBED was outperformed in DistillMOS by its predictive counterpart.
However, the converse is true on the mismatched test set. We see in DistillMOS that the generative method even outperforms the predictive baseline with larger $d$. Similarly, we observe in SI-SIR a relatively large difference of approximately 7 dB when $d$ is large enough. As SI-SIR is a metric that measures the amount of background noise removal, we conclude that the predictive baseline struggles to remove unseen impulsive noises\footnote{Audio examples in supplementary material.}.

We observe that increasing $d$ is important to ensure that the generative DB-BBED outperforms the predictive baseline on unseen data. We hypothesize that this is due to the following argument. Note that the stochastic process of BBED has no variance at Diffusion time-step $t=1$. This means, when taking only one step, thereby outputting the last frame from the Diffusion Buffer ($d=0$), then DB-BBED trained with the \ac{dp} is simply a direct mapping from $X_{\Tmax} \approx Y$ to $X_0=S$. Hence, it does not differ much from the predictive counterpart, which is also a direct mapping from $Y$ to $X_0$. Therefore, no improvement of the Diffusion Buffer over its predictive baseline can be expected when taking only a single step.

\section{Conclusion}
In this work, we build upon our existing work, the Diffusion Buffer \cite{diffusionbuffer}. The Diffusion Buffer is the first Diffusion model demonstrating that it performs Speech Enhancement online on consumer-grade GPUs.
As the original Diffusion Buffer from \cite{diffusionbuffer} outputs enhanced frames with a delay, the output frame has a certain look-ahead. However, the underlying 2D convolutional UNet does not align with this look-ahead constraint. This means that the network processes zeros due to zero-padding instead of processing only actual audio data. We showed that such a UNet has suboptimal performance when the algorithmic latency is low. We therefore designed a novel 2D convolutional UNet architecture whose receptive field is explicitly aligned with the look-ahead constraints of the Diffusion Buffer. In addition, we shifted from the Denoising Score Matching loss to the Data Prediction loss, which allows now to flexibly adjust the latency during streaming for the output frame, thereby trading enhancement performance for algorithmic latency. The newly developed Diffusion Buffer, equipped with the novel 2D convolutional UNet and trained on the data prediction loss, reduces latency to \qty{32}{}--\qty{176}{ms} opposed to \qty{320}{}--\qty{960}{ms}, while outperforming the original Diffusion Buffer. In addition, the newly developed Diffusion Buffer can now also perform live enhancement on an NVIDIA RTX 2080Ti, whereas previously a more powerful GPU (e.g., RTX 4080 Laptop) was required. Moreover, we showed experimentally that the generative Diffusion-Buffer generalizes better to unseen noise types than its predictive counterpart, that is, the underlying UNet trained in a predictive way.

\section{Acknowledgements}
Funded by the Deutsche Forschungsgemeinschaft (DFG, German Research Foundation) -- 545210893, 498394658; by the Federal Ministry for Economic Affairs and Climate Action (Bundesministerium für Wirtschaft und Klimaschutz), Zentrales Innovationsprogramm Mittelstand (ZIM), Germany, within the project FKZ KK5528802VW4; and by the German Federal Ministry of Research, Technology and Space (BMFTR) under grant agreement No. 01IS24072A (COMFORT).

The authors gratefully acknowledge the scientific support and HPC resources provided by the Erlangen National High Performance Computing Center (NHR@FAU) of the Friedrich-Alexander-Universität Erlangen-Nürnberg (FAU) under the NHR projects f101ac, f102ac. NHR funding is provided by federal and Bavarian state authorities. NHR@FAU hardware is partially funded by the German Research Foundation (DFG) – 440719683.

\bibliographystyle{IEEEtran}
\bibliography{bib_clean}

\begin{thebibliography}{10}
\providecommand{\url}[1]{#1}
\csname url@samestyle\endcsname
\providecommand{\newblock}{\relax}
\providecommand{\bibinfo}[2]{#2}
\providecommand{\BIBentrySTDinterwordspacing}{\spaceskip=0pt\relax}
\providecommand{\BIBentryALTinterwordstretchfactor}{4}
\providecommand{\BIBentryALTinterwordspacing}{\spaceskip=\fontdimen2\font plus
\BIBentryALTinterwordstretchfactor\fontdimen3\font minus \fontdimen4\font\relax}
\providecommand{\BIBforeignlanguage}[2]{{%
\expandafter\ifx\csname l@#1\endcsname\relax
\typeout{** WARNING: IEEEtran.bst: No hyphenation pattern has been}%
\typeout{** loaded for the language `#1'. Using the pattern for}%
\typeout{** the default language instead.}%
\else
\language=\csname l@#1\endcsname
\fi
#2}}
\providecommand{\BIBdecl}{\relax}
\BIBdecl

\bibitem{diffusionbuffer}
B.~Lay, R.~Makarov, and T.~Gerkmann, ``Diffusion buffer: Online diffusion-based speech enhancement with sub-second latency,'' \emph{ISCA Interspeech}, 2025.

\bibitem{defossez2020demucs}
A.~Defossez, G.~Synnaeve, and Y.~Adi, ``Real time speech enhancement in the waveform domain,'' in \emph{Interspeech}, 2020.

\bibitem{semamba}
R.~Chao, W.-H. Cheng, M.~L. Quatra, S.~M. Siniscalchi, C.-H.~H. Yang, S.-W. Fu, and Y.~Tsao, ``An investigation of incorporating mamba for speech enhancement,'' 2024, pp. 302--308.

\bibitem{richter_sgmse}
J.~Richter, S.~Welker, J.-M. Lemercier, B.~Lay, and T.~Gerkmann, ``Speech enhancement and dereverberation with diffusion-based generative models,'' \emph{IEEE Trans. on Audio, Speech, and Language Proc. (TASLP)}, 2023.

\bibitem{lu2021study}
Y.-J. Lu, Y.~Tsao, and S.~Watanabe, ``A study on speech enhancement based on diffusion probabilistic model,'' \emph{IEEE Asia-Pacific Signal and Inf. Proc. Assoc. Annual Summit and Conf. (APSIPA ASC)}, pp. 659--666, 2021.

\bibitem{lu2022conditional}
Y.-J. Lu, Z.-Q. Wang, S.~Watanabe, A.~Richard, C.~Yu, and Y.~Tsao, ``Conditional diffusion probabilistic model for speech enhancement,'' \emph{IEEE Int. Conf. on Acoustics, Speech and Signal Proc. (ICASSP)}, pp. 7402--7406, 2022.

\bibitem{welker2022speech}
S.~Welker, J.~Richter, and T.~Gerkmann, ``Speech enhancement with score-based generative models in the complex {STFT} domain,'' \emph{ISCA Interspeech}, pp. 2928--2932, 2022.

\bibitem{lay202interspeech}
B.~Lay, S.~Welker, J.~Richter, and T.~Gerkamnn, ``Reducing the prior mismatch of stochastic differential equations for diffusion-based speech enhancement,'' \emph{ISCA Interspeech}, 2023.

\bibitem{ho2020denoising}
J.~Ho, A.~Jain, and P.~Abbeel, ``Denoising diffusion probabilistic models,'' \emph{Advances in Neural Inf. Proc. Systems (NeurIPS)}, vol.~33, pp. 6840--6851, 2020.

\bibitem{jukic2024schr}
A.~Jukic, R.~Korostik, J.~Balam, and B.~Ginsburg, ``Schroedinger bridge for generative speech enhancement,'' \emph{IEEE Int. Conf. on Acoustics, Speech and Signal Proc. (ICASSP)}, 2024.

\bibitem{stochasticInterAlbergo}
M.~S. Albergo, M.~Goldstein, N.~M. Boffi, R.~Ranganath, and E.~Vanden-Eijnden, ``Stochastic interpolants with data-dependent couplings,'' 2024.

\bibitem{layCorrectingReverse}
B.~Lay, J.-M. Lemercier, J.~Richter, and T.~Gerkmann, ``Single and few-step diffusion for generative speech enhancement,'' \emph{IEEE Int. Conf. on Acoustics, Speech and Signal Proc. (ICASSP)}, 2024.

\bibitem{consistencymodel}
Y.~Song, P.~Dhariwal, M.~Chen, and I.~Sutskever, ``Consistency models,'' \emph{Int. Conf. on Machine Learning (ICML)}, 2023.

\bibitem{stream_diffusion}
C.~Li, S.~Cornell, S.~Watanabe, and Y.~Qian, ``Diffusion-based generative modeling with discriminative guidance for streamable speech enhancement,'' \emph{2024 IEEE Spoken Language Technology Workshop (SLT)}, pp. 333--340, 2024.

\bibitem{lemercier2023storm}
J.-M. Lemercier, J.~Richter, S.~Welker, and T.~Gerkmann, ``Storm: A diffusion-based stochastic regeneration model for speech enhancement and dereverberation,'' \emph{IEEE Trans. on Audio, Speech, and Language Proc. (TASLP)}, vol.~31, 2023.

\bibitem{lay_demo}
B.~Lay, R.~Makarov, and T.~Gerkmann, ``Real-time diffusion buffer for speech enhancement on a laptop,'' \emph{ISCA Interspeech}, 2025.

\bibitem{fifo}
J.~Kim, J.~Kang, J.~Choi, and B.~Han, ``Fifo-diffusion: Generating infinite videos from text without training,'' \emph{Advances in Neural Inf. Proc. Systems (NeurIPS)}, 2024.

\bibitem{rollingdiff}
D.~Ruhe, J.~Heek, T.~Salimans, and E.~Hoogeboom, ``Rolling diffusion models,'' \emph{Int. Conf. on Machine Learning (ICML)}, 2024.

\bibitem{Podell2023SDXLIL}
D.~Podell, Z.~English, K.~Lacey, A.~Blattmann, T.~Dockhorn, J.~Muller, J.~Penna, and R.~Rombach, ``Sdxl: Improving latent diffusion models for high-resolution image synthesis,'' \emph{International Conference on Learning Representations (ICLR)}, 2023.

\bibitem{kara_and_shreve}
I.~Karatzas and S.~E. Shreve, \emph{Brownian Motion and Stochastic Calculus}, 2nd~ed.\hskip 1em plus 0.5em minus 0.4em\relax Springer, 1996.

\bibitem{rudin}
W.~Rudin, \emph{Real and Complex Analysis}, 3rd~ed.\hskip 1em plus 0.5em minus 0.4em\relax McGraw-Hill, Inc., 1987.

\bibitem{sarkka2019sde}
S.~Särkkä and A.~Solin, \emph{Applied Stochastic Differential Equations}.\hskip 1em plus 0.5em minus 0.4em\relax {Cambridge University Press}, 2019.

\bibitem{anderson1982reverse}
B.~D. Anderson, ``Reverse-time diffusion equation models,'' \emph{Stochastic Processes and their Applications}, vol.~12, no.~3, pp. 313--326, 1982.

\bibitem{bender78:AMM}
C.~M. Bender and S.~A. Orszag, \emph{{Advanced Mathematical Methods for Scientists and Engineers}}.\hskip 1em plus 0.5em minus 0.4em\relax McGraw-Hill, 1978.

\bibitem{gradshteyn2007}
I.~S. Gradshteyn and I.~M. Ryzhik, \emph{Table of integrals, series, and products}, 7th~ed.\hskip 1em plus 0.5em minus 0.4em\relax Elsevier/Academic Press, Amsterdam, 2007.

\bibitem{song2021sde}
Y.~Song, J.~Sohl-Dickstein, D.~P. Kingma, A.~Kumar, S.~Ermon, and B.~Poole, ``Score-based generative modeling through stochastic differential equations,'' \emph{Int. Conf. on Learning Representations (ICLR)}, 2021.

\bibitem{zhang2019making}
R.~Zhang, ``Making convolutional networks shift-invariant again,'' \emph{Int. Conf. on Machine Learning (ICML)}, pp. 7324--7334, 2019.

\bibitem{vaswani2017attention}
A.~Vaswani, N.~Shazeer, N.~Parmar, J.~Uszkoreit, L.~Jones, A.~N. Gomez, {\L}.~Kaiser, and I.~Polosukhin, ``Attention is all you need,'' \emph{Advances in Neural Inf. Proc. Systems (NeurIPS)}, vol.~30, 2017.

\bibitem{karras2020analyzing}
T.~Karras, S.~Laine, M.~Aittala, J.~Hellsten, J.~Lehtinen, and T.~Aila, ``Analyzing and improving the image quality of {StyleGAN},'' \emph{IEEE/CVF Conf. on Computer Vision and Pattern Recognition (CVPR)}, pp. 8110--8119, 2020.

\bibitem{ojsp}
J.~Richter, S.~Welker, J.-M. Lemercier, B.~Lay, T.~Peer, and T.~Gerkmann, ``Causal diffusion models for generalized speech enhancement,'' \emph{IEEE Open Journal of Signal Processing}, vol.~5, pp. 780--789, 2024.

\bibitem{receptive_field}
W.~Luo, Y.~Li, R.~Urtasun, and R.~Zemel, ``Understanding the effective receptive field in deep convolutional neural networks,'' \emph{Advances in Neural Inf. Proc. Systems (NeurIPS)}, p. 4905–4913, 2016.

\bibitem{gerkmann2010empirical}
T.~Gerkmann and R.~Martin, ``Empirical distributions of {DFT-domain} speech coefficients based on estimated speech variances,'' \emph{Int. Workshop on Acoustic Echo and Noise Control}, 2010.

\bibitem{lay2024analysis}
B.~Lay and T.~Gerkmann, ``An analysis of the variance of diffusion-based speech enhancement,'' \emph{ISCA Interspeech}, 2024.

\bibitem{kingma2015adam}
D.~P. Kingma and J.~Ba, ``Adam: A method for stochastic optimization,'' \emph{Int. Conf. on Learning Representations (ICLR)}, 2015.

\bibitem{richter2024ears}
J.~Richter, Y.-C. Wu, S.~Krenn, S.~Welker, B.~Lay, S.~Watanabe, A.~Richard, and T.~Gerkmann, ``{EARS}: An anechoic fullband speech dataset benchmarked for speech enhancement and dereverberation,'' \emph{ISCA Interspeech}, 2024.

\bibitem{wham}
G.~Wichern, J.~Antognini, M.~Flynn, L.~Zhu, E.~McQuinn, D.~Crow, E.~Manilow, and J.~Le~Roux, ``Wham!: Extending speech separation to noisy environments,'' \emph{ISCA Interspeech}, 07 2019.

\bibitem{ANTONI2006282}
J.~Antoni, ``The spectral kurtosis: a useful tool for characterising non-stationary signals,'' \emph{Mechanical Systems and Signal Processing}, vol.~20, no.~2, pp. 282--307, 2006.

\bibitem{rixPerceptualEvaluationSpeech2001}
A.~Rix, J.~Beerends, M.~Hollier, and A.~Hekstra, ``Perceptual evaluation of speech quality ({{PESQ}}) - a new method for speech quality assessment of telephone networks and codecs,'' \emph{IEEE Int. Conf. on Acoustics, Speech and Signal Proc. (ICASSP)}, vol.~2, pp. 749--752, 2001.

\bibitem{jensen2016algorithm}
J.~Jensen and C.~H. Taal, ``An algorithm for predicting the intelligibility of speech masked by modulated noise maskers,'' \emph{IEEE Trans. on Audio, Speech, and Language Proc. (TASLP)}, vol.~24, no.~11, pp. 2009--2022, 2016.

\bibitem{leroux2018sdr}
J.~Le~Roux, S.~Wisdom, H.~Erdogan, and J.~R. Hershey, ``{SDR}--half-baked or well done?'' \emph{IEEE Int. Conf. on Acoustics, Speech and Signal Proc. (ICASSP)}, pp. 626--630, 2019.

\bibitem{stahl2025distillation}
B.~Stahl and H.~Gamper, ``Distillation and pruning for scalable self-supervised representation-based speech quality assessment,'' \emph{IEEE Int. Conf. on Acoustics, Speech and Signal Proc. (ICASSP)}, 2025.

\bibitem{exploretradeoffsWilson}
K.~Wilson, M.~Chinen, J.~Thorpe, B.~Patton, J.~Hershey, R.~Saurous, J.~Skoglund, and R.~Lyon, ``Exploring tradeoffs in models for low-latency speech enhancement,'' \emph{International Workshop on Acoustic Signal Enhancement (IWAENC)}, 11 2018.

\end{thebibliography}

\section{Biography Section}
\begin{IEEEbiography}[{\includegraphics[width=1in,height=1.25in,clip,keepaspectratio]{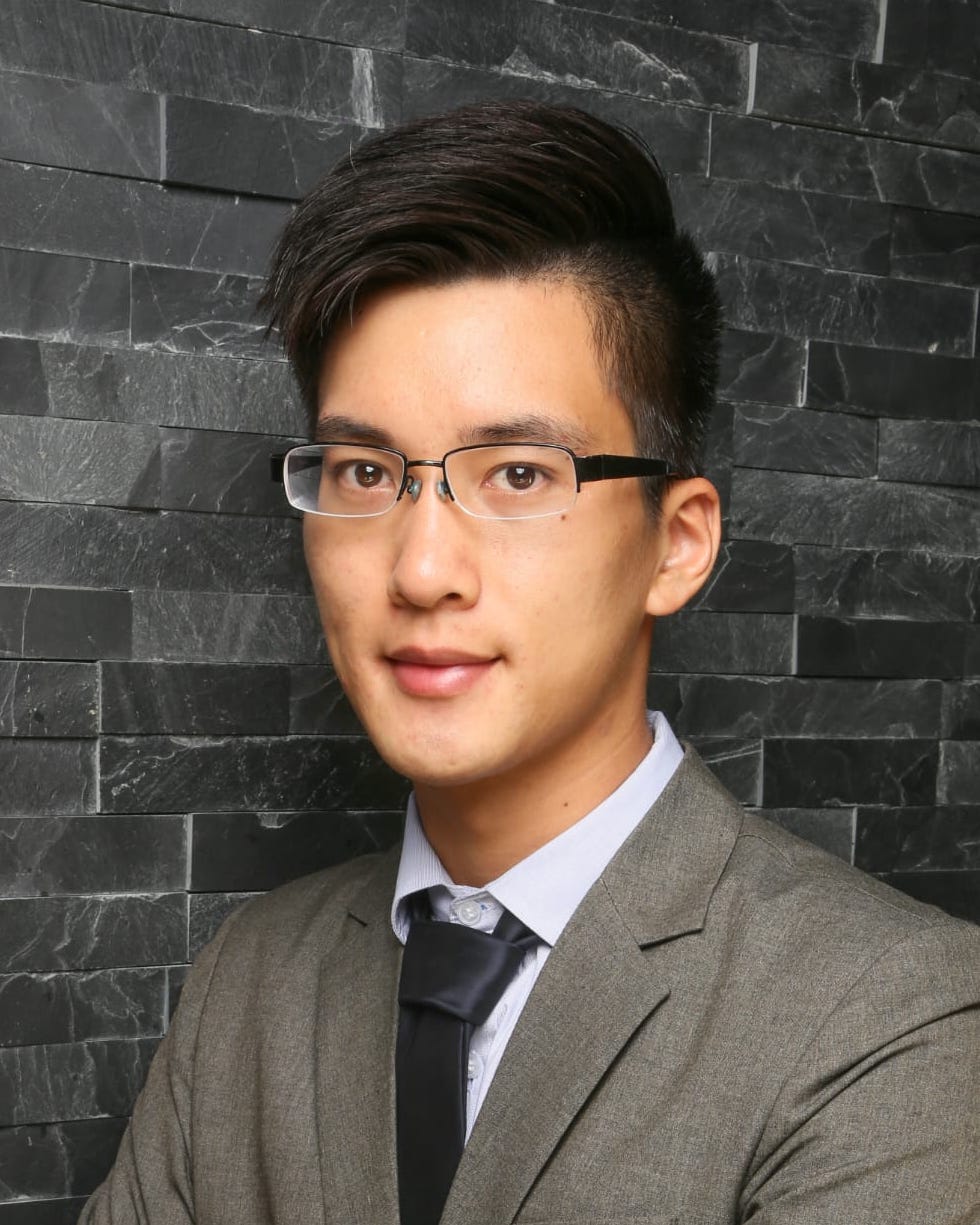}}]{Bunlong Lay}
obtained a B.Sc. and M.Sc. in Mathematics in 2015 and 2017 from the University of Bonn, Germany. He subsequently joined the research institute 
Fraunhofer FKIE in Wachtberg Germany from 2018 until 2021, where he focused on research in the field of radar signal processing. In 2021 he started his Ph.D. at the University of Hamburg. Currently researching Diffusion-based models for Speech Enhancement for real-time applications. He received the VDE ITG award 2024.
\end{IEEEbiography}

\begin{IEEEbiography}[{\includegraphics[width=1in,height=1.25in,clip,keepaspectratio]{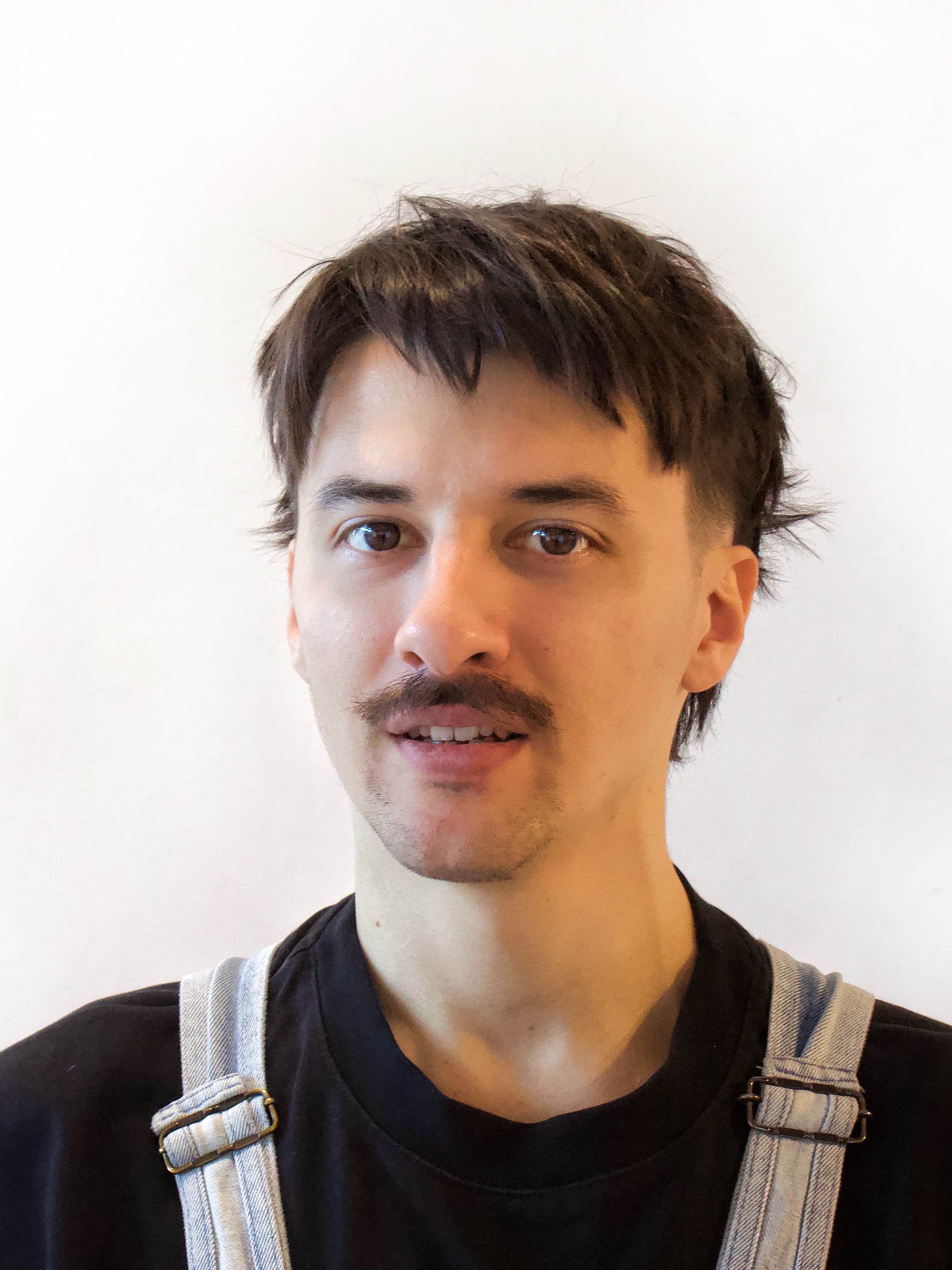}}]{Rostislav Makarov}
received the B.Sc. in Laser Physics (2017) and the M.Sc. in Information Systems and Technology (2019) from ITMO University, Saint Petersburg, Russia. From 2019 to 2025, he was a Machine Learning Research Engineer at ID R\&D, researching production-grade systems for speaker recognition, anti-spoofing, and speech processing. Since 2025, he has been pursuing his Ph.D. at the Universität Hamburg, focusing on Diffusion-based generative models for speech enhancement. %
\end{IEEEbiography}

\begin{IEEEbiography}[{\includegraphics[width=1in,height=1.25in,clip,keepaspectratio]{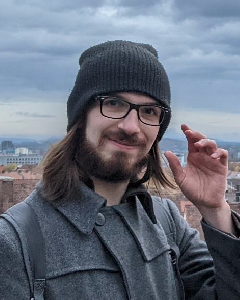}}]{Simon Welker}
received a B.Sc. in Computing in Science (2019) and M.Sc. in Bioinformatics (2021) from Universität Hamburg, Germany. He is currently a fourth-year PhD student in the labs of Prof. Timo Gerkmann (Signal Processing, Universität Hamburg) and Prof. Henry N. Chapman (Center for Free-Electron Laser Science, DESY, Hamburg), researching machine learning techniques for solving inverse problems that arise in speech processing and X-ray imaging. He received the VDE ITG award 2024.
\end{IEEEbiography}

\begin{IEEEbiography}[{\includegraphics[width=1in,height=1.25in,clip,keepaspectratio]{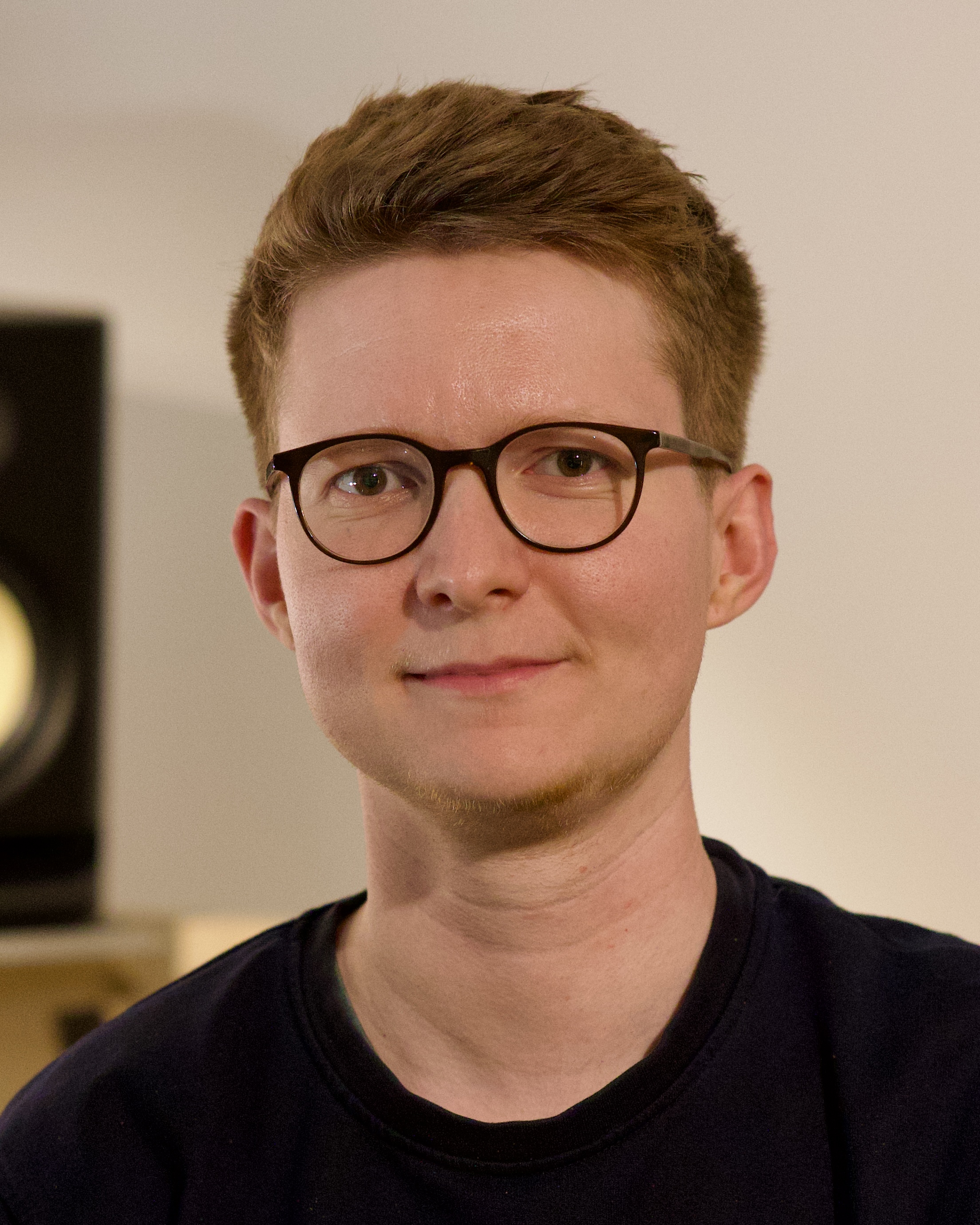}}]{Maris Hillemann}
obtained a B.Sc. in Computer Science in 2024 from the University of Hamburg, Germany. He is currently a master's student in Computer Science at the University of Hamburg and a student assistant in the Signal Processing group of Prof. Timo Gerkmann.
\end{IEEEbiography}

\begin{IEEEbiography}[{\includegraphics[width=1in,height=1.25in,clip,keepaspectratio]{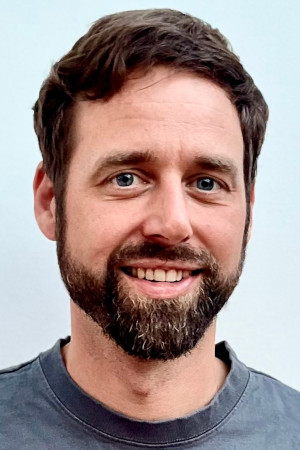}}]{Timo Gerkmann}
(S’08–M’10–SM’15) is a professor for Signal Processing at the Universität Hamburg, Germany. He has previously held positions at Technicolor Research \& Innovation in Germany, the University of Oldenburg in Germany, KTH Royal Institute of Technology in Sweden, Ruhr-Universität Bochum in Germany, and Siemens Corporate Research in Princeton, NJ, USA. His main research interests are on statistical signal processing and machine learning for speech and audio applied to communication devices, hearing instruments, audio-visual media, and human-machine interfaces. Timo Gerkmann served as member of the IEEE Signal Processing Society Technical Committee on Audio and Acoustic Signal Processing and is currently a Senior Area Editor of the IEEE/ACM Transactions on Audio, Speech and Language Processing. He received the VDE ITG award 2022.
\end{IEEEbiography}

\vfill

\end{document}